\newcommand{\diracslash}[1]{#1\llap{/\kern2pt}}

\newcommand{\be}{\begin{equation}}
\newcommand{\ee}{\end{equation}}
\newcommand{\bea}{\begin{eqnarray}}
\newcommand{\eea}{\end{eqnarray}}
\newcommand{\ba}[1]{\begin{array}{#1}}
\newcommand{\ea}{\end{array}}

\newcommand{\bt}{\begin{tabular}}
\newcommand{\et}{\end{tabular}}

\newcommand{\beas}{\begin{eqnarray*}}
\newcommand{\eeas}{\end{eqnarray*}}

\RequirePackage{lineno}

\documentclass[preprint,prd,aps,floats,nofootinbib,floatfix]
 {revtex4-2}

\DeclareSymbolFont{rsfs}{U}{rsfs}{m}{n}
\DeclareSymbolFontAlphabet{\mathrsfs}{rsfs}

\usepackage{amsmath}
\usepackage{graphicx}
\usepackage[mathscr]{eucal}
\usepackage{multirow}
\usepackage{graphicx,epstopdf}
\usepackage{graphicx}
\usepackage{subcaption}
\usepackage{cancel}

\usepackage{setspace}
\usepackage[utf8]{inputenc}
\usepackage{csquotes}
\usepackage[english]{babel}
\usepackage{hyperref}
\usepackage{upgreek}
\hypersetup{
    colorlinks=true,
    linkcolor=blue,
    filecolor=magenta,     
    citecolor=blue, 
    urlcolor=blue,
}
\usepackage{cleveref}
\usepackage{caption}

\begin{document}
\setstretch{1.5}
%\begin{linenumbers}
%\begin{linenumbers}
\title{Kaons and antikaons in isospin asymmetric dense resonance matter at finite temperature} 
%%%%%%%%%%%%%%%%%%%%%%%%%%%%%%%%%%%%%%%%%%%%%%%%%%%%%%%%%%%%%%%%%
\author{Manpreet Kaur}
\email{ranapreeti803@gmail.com}
\affiliation{Department of Physics, Dr. B R Ambedkar National Institute of Technology Jalandhar, 
	Jalandhar -- 144008, Punjab, India}

\author{Arvind Kumar}
\email{kumara@nitj.ac.in}
\affiliation{Department of Physics, Dr. B R Ambedkar National Institute of Technology Jalandhar, 
	Jalandhar -- 144008, Punjab, India}

\def\be{\begin{equation}}
\def\ee{\end{equation}}
\def\bearr{\begin{eqnarray}}
\def\eearr{\end{eqnarray}}
\def\zbf#1{{\bf {#1}}}
\def\bfm#1{\mbox{\boldmath $#1$}}
\def\hf{\frac{1}{2}}
\def\kp{\zbf k+\frac{\zbf q}{2}}
\def\km{-\zbf k+\frac{\zbf q}{2}}
\def\hwo{\hat\omega_1}
\def\hwt{\hat\omega_2}

\begin{abstract}
We study the in-medium properties of kaons and antikaons in isospin asymmetric hot and dense resonance matter within the chiral SU(3) hadronic mean field model.     
Along with nucleons and hyperons, the interactions of $K$  and $\bar K$ mesons with all decuplet baryons ($\Delta^{++,+,0,-}, \Sigma^{*\pm,0},\Xi^{*0,-}, \Omega^{-}$) are explicitly considered in the dispersion relations.
The properties of mesons in the chiral SU(3) model are modified at finite density and temperature of asymmetric resonance matter through the exchange of scalar fields $\sigma, \zeta$ and $\delta$ and the vector fields $\omega, \rho$ and $\phi$. 
 The presence of  resonance baryons
 in the medium at finite temperature is observed to modify significantly
 the effective masses of $K$ and $\bar{K}$ mesons. We also calculated the optical potentials of kaons and antikaons as a function of momentum in resonance matter.
  The present study of in-medium masses and optical potentials of kaons and antikaons will be important for understanding the experimental observables from the heavy-ion collision experiments where hot and dense matter may be produced. Our results indicate that when resonance baryons are present within the medium at finite baryonic density, the mass reduction of kaons and antikaons becomes more pronounced as the temperature of the medium increases from zero to 100 and 150 MeV. The study of the optical potentials of kaons and antikaons reveals a stronger correlation with strangeness fraction compared to isospin asymmetry.

\end{abstract}

\maketitle

\newpage

\section{Introduction}
\label{intro}
The study of
% thermodynamic
the properties of hadrons in isospin asymmetric dense matter at finite density and temperature is a significant topic of research that finds relevance in analyzing the experimental outcomes from different  relativistic heavy-ion collision (HIC) experiments and also in exploring the structure of compact objects \cite{hayano2010hadron,Holzenkamp:1989tq,Haidenbauer:2018gvg,Rafelski:2000by,Pushkina:2004wa,Knospe:2017oin}.
% Hadrons have a large mass relative to their quark constituents because of the spontaneous and explicit breaking of the chiral symmetry in QCD. 
% % This breaking results in a vacuum populated by quark condensates and in-medium mass modification in the hadronic medium. 
% The modifications of the hadron properties are directly relevant to the experimental observables of heavy-ion collision experiments. For example, the CERES \cite{CERES:1995vll}, HELIOS \cite{masera1995dimuon}, and DLS \cite{porter1997dielectron}  collaborations aim to identify modifications of hadron properties in hot and dense matter via the measurement of dilepton production in heavy ion collisions \cite{bratkovskaya1997dilepton, mishra2001vector}.
Depending upon the center of mass energy of the colliding beams, different HIC experimental facilities are dedicated to 
explore the regimes of quantum chromodynamics (QCD) phase diagram at different temperatures and baryon densities.
The HIC experiments such as the relativistic heavy ion collider (RHIC) \cite{Harrison:2003sb,friman2011cbm,kumar2013star,odyniec2010rhic} and the large hadron collider (LHC)  \cite{evans2007large,DiNezza:2024ctw} have explored the region of QCD phase diagram at high temperature and almost baryon-free regime, i.e., low baryon density and high temperature phase, which might have existed during the early evolution of the universe.
On this side, the phase transition from the hadronic
to QGP phase has been predicted as a simple crossover.
Exploring the dense matter at finite baryon densities and low/moderate temperatures is the aim of  beam energy scan programs, BES-I and BES-II, of RHIC \cite{Odyniec:2019kfh,Liu:2022cqh, Liu:2022wme}, the compressed baryonic matter (CBM) experiment of Facility for Antiproton and Ion Research (FAIR) at GSI \cite{senger2012compressed,rapp2010charmonium}, Japan Proton Accelerator Research Complex (J-PARC Japan) \cite{Sawada:2007gy,Hotchi:2017krx} and Nuclotron-based Ion Collider Facility (NICA) at Dubna, Russia \cite{Kekelidze:2017ghu,Kekelidze:2017tgp}.

%, aim to understand quantum chromodynamics (QCD) phase diagram at different regimes of density and temperature. RHIC and LHC investigate the properties of hadronic matter at high temperature and low baryonic density conditions, whereas heavy ion collision experiments at other experimental facilities intend to explore it at moderate temperatures and high baryonic densities \cite{chattopadhyay2015present,chhabra2017medium}.

\par
 In HICs, two beams of heavy ions collide head-on in a controlled environment, melting them and releasing their constituent quarks and gluons. The perturbative techniques can be applied in the asymptotic regime, where quarks and gluons are free. On the other hand, as the temperature decreases with the expansion of the collision zone, hadronization takes place. In this phase, quarks and gluons are confined within the hadrons, which means the value of the coupling constant is large, and hence, the perturbative theory is no longer applicable in a low-energy regime, and one requires the non-perturbative approaches. 
% The measurement of hadronic resonance provides a deeper understanding of fundamental aspects of QCD, like confinement and chiral symmetry breaking.
  Lattice QCD is a simulation approach solved on grid points of space-time to study the QCD phase diagram at high temperature and  baryon-free regime \cite{wilson1974confinement}.
%This framework has revealed the quenched QCD properties at finite density and temperature.  
However, as the baryon density increases, lattice QCD becomes inapplicable because of the sign problem of the fermion determinant, which makes oscillations more pronounced, resulting in increased computational expenses and reduced productivity \cite{alexandru2022complex,bloch2009random,fukushima2007model,muroya2003lattice,takaishi2004hadronic}. 
 To explore the hadronic matter beyond the lattice QCD regime, various theoretical frameworks have been proposed, such as the coupled channel approach \cite{ramos2000properties,Schaffner-Bielich:1999fyk,Lutz:1997wt,tolos2004properties,tolos2006d, tolos2008open, Wang:2024hwu,Liu:2023gla,Wilson:2023hzu}, quark meson coupling (QMC) model \cite{tsushima1999charmed, saito1994quark,Tsushima:1997cu}, Polyakov quark meson (PQM) model \cite{Schaefer:2007pw,Herbst:2010rf,  Stiele:2013pma}, Polyakov loop extended NJL (PNJL) model \cite{kashiwa2008critical,Fukushima:2003fw}, chiral perturbation theory (ChPT) \cite{pich1995chiral,scherer2003introduction,Waas:1996tw}, QCD sum rules \cite{hatsuda1992qcd,hayano2010hadron,kumar2011d}, chiral SU(3) quark mean field (CQMF) model \cite{Zhang:1997ny,Wang:2004wja} and chiral SU(3) model \cite{Papazoglou:1998vr,Mishra:2009bp,Dexheimer:2008ax,Kumar:2018pqs,kumar2020phi,chahal2024phi,Zschiesche:2003qq,Cruz-Camacho:2024odu}. 
 Using these different non-perturbative approaches, the impact of finite baryon density and temperature has been analyzed on the properties, such as in-medium masses, decay width, decay constants, spectral properties etc., of the different mesons. In most of available studies, while exploring the in-medium masses and optical potentials of different mesons, the baryonic matter considered is either nuclear matter (composed of neutrons and protons) or strange matter
 (nucleons and hyperons, i.e., baryon octet). The properties of light vector mesons have been studied using the QCD sum rules \cite{hayano2010hadron,kumar2011d,Hatsuda:1996xt}, coupled channel approach \cite{ramos2000properties,Schaffner-Bielich:1999fyk}, and chiral SU(3) model \cite{Kumar:2018pqs,kumar2020phi,chahal2024phi,Zschiesche:2003qq} in nuclear matter. 
 In Ref. \cite{Mishra:2014rha,kumar2020phi}, the in-medium masses of vector mesons are studied in strange hadronic medium. The in-medium masses and optical potentials of kaons \cite{Mishra:2006wy,Mishra:2008dj,Mishra:2008kg,Tolos:2000fj,Tolos:2001js,Cobos-Martinez:2017vtr,Gifari:2024ssz}, open charm $D$ mesons \cite{Tolos:2009nn,Kumar:2010gb,kumar2011d,Suzuki:2017spp}, $B$ mesons \cite{Pathak:2014nfa,Chhabra:2016uom}, charmonium \cite{S:2021qkj,De:2022gse,Kumar:2010hs} and bottomonium \cite{Mishra:2014gea,Zeminiani:2020aho,Waas:1996tw} have also been investigated in the nuclear and hyperonic matter using various approaches at finite density and temperature of the medium.
 
 Few studies which explore the properties of dense resonance matter and also, how the properties of resonance baryons are modified in nuclear and strange matter are available in the literature.
For example,
the impact of finite temperature on the masses of decuplet baryons is investigated using the heavy baryon chiral perturbation theory \cite{Bedaque:1995pa}, NJL and PNJL model \cite{Torres-Rincon:2015rma} and thermal QCD sum rules \cite{Xu:2015jxa,Azizi:2016ddw}.
 Using the QCD sum rules, the in-medium masses of decuplet baryons, $\Delta, \Sigma^*, \Xi^*, \Omega^{-}$ have been studied in symmetric nuclear  matter at zero temperature \cite{Azizi:2016hbr}.
 Employing the  real time formalism of thermal field theory the self-energy of $\Delta$ resonance baryons has been studied at finite temperature and density \cite{Ghosh:2016hln}.
The in-medium mass and magnetic moment of decuplet baryons have been studied in nuclear and hyperons medium by using the chiral SU(3) quark mean field model \cite{Kumar:2023owb, Ryu:2008st, Singh:2017mxj, Singh:2020nwp}.
   In Ref. \cite{Zschiesche:2000ew}, chiral SU(3) model is used to investigate the properties of dense resonance matter at zero temperature. The relativistic mean-field models including $\Delta$ resonances
  have been used to study the equation of state of dense matter \cite{Lavagno:2010ah,Kolomeitsev:2016ptu}. The properties of heavy baryon resonances have also been studied in the nuclear matter \cite{Lenske:2018bvq,Singh:2017mxj}. The impact of resonance baryons, mostly $\Delta$ resonances, has been explored using different approaches on the mass-radius relations of compact stars \cite{Glendenning:1982nc, Oliveira:2019xni, Raduta:2020fdn, Dexheimer:2021sxs} and is currently an active topic of research.
 
 Keeping in view the above discussed literature, the studies which investigate thoroughly the properties of mesons, for example their effective masses, in a medium composed of spin$-\frac{3}{2}$ resonance baryons, in addition to the  spin$-\frac{1}{2}$ octet baryons and having finite temperature,  still need considerable attentions.
 Therefore, in the present work we aim to investigate the in-medium masses and optical potentials of kaons, $K(K^{+},K^{0})$, and antikaons, $\bar{K} (K^{-}, \bar{K^0})$, in isospin asymmetric matter, which consists of nucleons, hyperons and spin$-\frac{3}{2}$ decuplet baryons.
 We shall use the chiral SU(3) hadronic mean field model for our present investigation \cite{Papazoglou:1998vr}. The chiral SU(3) model has been used in literature to study 
 the properties of kaons and antikaons in nuclear \cite{Mishra:2006wy,Mishra:2008dj} and strange matter \cite{kumar2020phi} at finite density and temperature. The model has also been generalized to SU(4) and SU(5)
 sectors to investigate the properties of $D$ \cite{Tolos:2009nn,Kumar:2010gb,kumar2011d} and $B$ mesons \cite{Pathak:2014nfa}. 
 %The chiral SU(3) model has also been used in combination with QCD sum rules to study the in-medium properties of different mesons \cite{bibid}.
  The model has also been applied 
 to investigate the impact of strong external magnetic field 
on the properties of mesons \cite{ReddyP:2017tgo,Kumar:2020kng,Kumar:2019axp,Mishra:2023uhx}.

In the present work, to explore the properties of $K(K^{+},K^{0})$ and  $\bar{K} (K^{-}, \bar{K^0})$ mesons in dense resonance matter using the framework of the chiral SU(3) model,  we shall find the explicit interactions of these mesons with different resonance baryons and shall find the effective masses and optical potentials at different densities and temperature. We shall explore the impact  
of varying the isospin asymmetry and strangeness fraction in the resonance medium on the properties of kaons and antikaons. 
 In the literature, the coupled channel approach has also been used to study the properties antikaons in the nuclear matter with aim to explore the dynamics of $\Lambda(1405)$ resonance baryon \cite{Koch:1994mj}. Such coupled channel effects are not considered in our present work. Also, the properties of antikaons in dense resonance matter will also be important for the physics of compact stars in connection with antikaon condensation. The studies related to compact stars, considering charge neutral and beta-equilibrated matter will be carried out in future work.
 Within the chiral SU(3) model the properties of baryons are modified through the exchange of
non-strange scalar isoscalar field $\sigma$, strange scalar field $\zeta$, scalar isovector field $\delta$ and also the vector fields $\omega, \rho$ and $\phi$.
The scalar isovector field $\delta$ and the vector-isovector field $\rho$ account for the finite isospin asymmetry of the medium.

This paper is organized as follows: In Sec.~\ref{method_chiral} we present the details of the chiral SU(3) hadronic mean field model  used to  study the properties of kaons and antikaons. In Sec.~\ref{subsec2.2} the interaction Lagrangian densities and self-energies for kaons and antikaons in resonance matter are presented. In Sec.~\ref{results} we give the results and discussions and finally, the work is summarized   in Sec.~\ref{summary}.

\section{\label{method} Methodology}
	% QCD remains unsolvable when dealing with low energy regimes and finite baryon densities. Thus, exploring another method to describe the properties of hadron matter becomes necessary. The
 % effective models, which focus on the relevant degree of freedom, are only solvable and carry all the information from all the theories. 
 We use the chiral SU(3) model to obtain the effective masses and optical potentials of $K$ and $\bar{K}$ mesons in dense resonance matter at finite temperature, considering finite 
 isospin asymmetry.
 In the following Sec.~\ref{method_chiral}, we discuss the details of the chiral SU(3) model and shall obtain the density and temperature dependent values of the scalar fields, $\sigma, \zeta$ and $\delta$ and vector fields, $\omega, \rho$ and $\phi$.  As we shall see below, the effective masses of baryons are expressed in terms of scalar fields, whereas the vector fields contribute through the effective chemical potentials of baryons. 
 The  Lagrangian density, describing the interactions of pseudoscalar $K$ and $\bar{K}$ mesons with baryons octet as well as the decuplet resonances is described in Sec.~\ref{subsec2.2}.
 
%  on scalar and vector fields. The following section provides a brief description of this model. 
 
\subsection{ The hadronic chiral  SU(3) model} \label{method_chiral}
The chiral SU(3) model is an effective theory applied in the non-perturbative regime of QCD to describe the hadron-hadron interactions. It is based on the nonlinear realization \cite{bardeen1969some} and broken scale invariance properties of chiral symmetry \cite{Papazoglou:1998vr, weinberg1968nonlinear}.
The effective Lagrangian density  of the chiral SU(3) model is written as
\begin{equation}
\label{gen_L}
\mathcal{L} 
  = \mathcal{L}_{\text{kin}} + \sum_{M=P,X,V,A} \mathcal{L}_{{BM}} +
	\mathcal{L}_{0}+ \mathcal{L}_{\text{vec}} + \mathcal{L}_{\text{SB}}.  
\end{equation}
\par
The term $\mathcal{L}_{kin}$ includes the kinetic energy terms for different particles. In general, we write
\begin{align}
\mathcal{L}_{\text {kin }} &=i \operatorname{Tr} \bar{B} \gamma_{\mu} D^{\mu} B+\frac{1}{2} \operatorname{Tr} D_{\mu} X D^{\mu} X + \operatorname{Tr}\left(u_{\mu} X u^{\mu} X + X u_{\mu} u^{\mu} X\right)
	\quad+\frac{1}{2} \operatorname{Tr} D_{\mu} Y D^{\mu} Y \nonumber \\
	+&\frac{1}{2} D_{\mu} \chi D^{\mu} \chi-\frac{1}{4} \operatorname{Tr}\left(V_{\mu \nu} V^{\mu \nu}\right)-\frac{1}{4} \operatorname{Tr}\left(\mathcal{A}_{\mu \nu} \mathcal{A}^{\mu \nu}\right).  
	\label{Eq_kinetic_2}
\end{align}
 The kinetic energy term for the baryon octet is given by the first term of the above equation.  
In this term, covariant derivative $D_{\mu}$  is defined as $D_\mu B = \partial_\mu B + i \left[\Gamma_\mu, B\right]$, with
$\Gamma_{\mu}=-\frac{i}{4}\left[u^{\dagger} \partial_{\mu} u-\partial_{\mu} u^{\dagger} u+\right.$ $\left.u \partial_{\mu} u^{\dagger}-\partial_{\mu} u u^{\dagger}\right]$.
Also, $u=\exp \left[\frac{i}{\sigma_{0}} \pi^{a} \lambda^{a} \gamma_{5}\right] = \exp\left[i P/ \sqrt{2} \sigma_0\right]$, is the unitary transformation operator defined in terms of pseudoscalar meson matrix $P$ \cite{Papazoglou:1998vr}.
As we shall see in Sec.~\ref{subsec2.2},
the interactions of pseudoscalar $K$ and $\bar{K}$ mesons with nucleons and hyperons (Weinberg Tomozawa term) will be obtained from the first term of Eq.~(\ref{Eq_kinetic_2}). 
%The pseudoscalar mesons come into the picture through the unitary transformation operator $u$ \cite{Papazoglou:1998vr}.
 The second term of Eq.~(\ref{Eq_kinetic_2}) is the kinetic energy for the scalar mesons whereas the third term is for pseudoscalar mesons. Here, $X$ denotes the scalar meson matrix and $u_{\mu} =-\frac{i}{2} \left[u^{\dagger}(\partial_{\mu}u) 
-u (\partial_{\mu}u^\dagger) \right]$.
The fourth and fifth terms represent the kinetic energy terms for the pseudoscalar singlet $Y$ and the dilaton field $\chi$, respectively. The term having field tensors, $ V_{\mu \nu}$, represents the kinetic term of spin-1 vector mesons whereas the last term with field tensor, $ \mathcal{A}_{\mu \nu}$, is for the axial vector mesons.

The second term of Eq.~(\ref{gen_L}) gives the interactions of baryons with different mesons. The mass term for baryons is obtained through their interactions with the scalar mesons. Explicitly, the interaction term
of baryons with scalars, $\sigma, \zeta$ and $\delta$, and vectors, $\omega, \rho$ and $\phi$, is written as
\begin{equation}
\mathcal{L}_{BW}=-\sum_i \bar{\psi}_i[m_i^*+g_{\omega i} \gamma_0 \omega+g_{\rho i} \gamma_0 I_{3i} \rho+g_{\phi i} \gamma_0 \phi] \psi_i.
\label{Eq_Lag_BW}
\end{equation}
In the above equation, the summation is over all the octet and decuplet baryons considered in the present work.
The field $\psi_i$ is for the spin$-\frac{1}{2}$ octet $(p,n,\Lambda,\Sigma^{\pm,0}, \Xi^{-,0})$ as
well as  spin$-\frac{3}{2}$ decuplet
($\Delta^{++,+,0,-}, \Sigma^{*\pm,0},\Xi^{*0,-}, \Omega^{-}$) baryons.
Though the spin$-\frac{3}{2}$ baryons are described by the  Rarita-Schwinger Lagrangian density, their equations of motion can be written as Dirac equation along with some supplementary conditions \cite{dePaoli:2012eq}. Considering the form of Lagrangian density for spin$-\frac{3}{2}$ as given by
Eq.~(\ref{Eq_Lag_BW}), the properties of neutron stars in the presence of decuplet baryons have been studied using different relativistic mean field models \cite{Marquez:2022gmu,Sedrakian:2022ata}. 
 %where summation over $i$ for all baryons (octet and decuplet baryons). 
In Eq.~(\ref{Eq_Lag_BW}) the in-medium baryon mass, $m_i^*$, is given as
\begin{equation}
 m_i^*=-\left(g_{\sigma i} \sigma+g_{\zeta i} \zeta+g_{\delta i} I_{3i} \delta\right)+ m_{i0},
 \label{mass_B}
\end{equation}
 where  $g_{\sigma i}$, $g_{\zeta i}$, and $g_{\delta i}$ are couplings of scalar fields $\sigma$, $\zeta$, and $\delta$, respectively, with different baryons. The additional mass term $m_{i0}$ is determined to fit the vacuum masses of baryons.
   % , i.e., $M_N=939$ MeV, $M_\Lambda=1115$ MeV, $M_\Sigma=1193$ MeV, $M_\Xi=1315$ MeV, $M_\Delta=1232$ MeV, $M_{\Sigma^*}=1385$ MeV, $M_{\Xi^*}=1585$ MeV, $M_\Omega=1672$ MeV.
The third term, $\mathcal{L}_{0}$, of Eq.~(\ref{gen_L}) represents the spontaneous chiral symmetry breaking and contains the self-interaction terms of scalar mesons. This term  is given by
\begin{eqnarray}
\label{eq_L0}
	&&\mathcal{L}_0=-\frac{1}{2} k_0 \chi^2\left(\sigma^2+\delta^2+\zeta^2\right)+k_1\left(\sigma^2+\delta^2+\zeta^2\right)^2 +k_2\left(\frac{\sigma^4}{2}+\frac{\delta^4}{2}+3 \sigma^2 \delta^2+\zeta^4\right)\nonumber\\
		&& +k_3 \chi \zeta\left(\sigma^2-\delta^2\right)-k_4 \chi^4-\frac{1}{4} \chi^4 \ln \frac{\chi^4}{\chi_0^4}+\frac{d}{3} \chi^4 \ln \left(\left(\frac{\left(\sigma^2-\delta^2\right) \zeta}{\sigma_0^2 \zeta_0}\right)\left(\frac{\chi}{\chi_0}\right)^3\right) ,
\end{eqnarray}
 where $\sigma_0, \zeta_0$, and $\chi_0$ represent the respective vacuum values of $\sigma, \zeta$, and $\chi$ fields. 
 %The fourth term, \(\mathcal{L}_{\text{vec}}\), generates the
 The mass term for the vector mesons as well as their self interactions are introduced through the relation (fourth term in Eq.~(\ref{gen_L}))
 \begin{eqnarray}
 \label{eq_Lvec}
 \mathcal{L}_{\text {vec }}=\frac{1}{2}\left(m_\omega^2 \omega^2+m_\rho^2 \rho^2+m_\phi^2 \phi^2\right) \frac{\chi^2}{\chi_0^2}+g_4\left(\omega^4+6 \omega^2 \rho^2+\rho^4+2 \phi^4\right) .
 \end{eqnarray} 
 The last term, $\mathcal{L}_{SB}$, of Eq.~(\ref{gen_L}) describes the explicit chiral symmetry breaking (eliminates the Goldstone bosons)
 of the chiral SU(3) model and its general expression is  given by
\begin{equation}
\label{Eq_SB_chiral1}
 {\mathcal L}_{SB}  =-\frac{1}{2} Tr A_p \left(uXu+u^{\dagger}Xu^{\dagger}\right) , 
\end{equation}
where $A_p=\frac{1}{\sqrt{2}}~ {\mathrm{diag}} (m_{\pi}^2 f_{\pi},
 m_\pi^2 f_\pi,
  2 m_K^2 f_K 
-m_{\pi}^2 f_\pi)$. Expanding the operator, $u = \exp\left[i P/ \sqrt{2} \sigma_0\right] = 1 + i P/ \sqrt{2} \sigma_0$ and keeping the first term in the expansion will lead to the following expression for symmetry breaking term
\begin{eqnarray}
\label{eq_LSB}
\mathcal{L}_{S B}=-\left(\frac{\chi}{\chi_0}\right)^2\left[m_\pi^2 f_\pi \sigma+\left(\sqrt{2} m_K^2 f_K-\frac{1}{\sqrt{2}} m_\pi^2 f_\pi\right) \zeta\right] .
\end{eqnarray}
 In the above equation, $f_\pi$ and $f_K$ are the decay constants of pions and kaon, respectively, whereas, $m_\pi$ and $m_K$ are their masses in
 the free space. The second term in the expansion of $u$  gives the interactions of pseudoscalar mesons with scalar mesons ($X$ is the scalar meson matrix), as discussed in the next section.
% 
% 
% \begin{equation}
% \label{33}
%  {\mathcal L}_{mass}  =-\frac{1}{2} Tr A_p \left(uXu+u^{\dagger}Xu^{\dagger}\right) , 
% \end{equation}
% where $X$ is the scalar meson multiplet and $A_{p}$ given as,
% \begin{eqnarray}
% A_p&=&1/\sqrt{2} {\mathrm{diag}} (m_{\pi}^2 f_{\pi},
%  m_\pi^2 f_\pi,
%   2 m_K^2 f_K 
% -m_{\pi}^2 f_\pi) .
% \end{eqnarray}
 
% With $f_\pi$=93.3 MeV and  $f_K$=122 MeV, we calculate  $\sigma_0$=93.3 MeV and  $\zeta_0$=106.7 MeV respectively .
To study the properties of baryons and mesons at finite temperature and density of the medium, we first write the partition function of the grand canonical ensemble given as \cite{Zschiesche:1999gf} 
\begin{eqnarray}
\mathcal{Z}={Tr} \exp \left[-\beta\left(\hat{\mathcal{H}}-\sum_i \mu_i \hat{\mathcal{N}}_i\right)\right] ,
\end{eqnarray}
where $\beta=1 / T$, $\hat{\mathcal{H}}$ and $\hat{\mathcal{N}_i}$ are the operators of Hamiltonian density and number density, respectively, whereas $\mu_i$ represents the chemical potential of baryon.
\par
At given temperature $T$, the thermodynamic potential, $\Omega$ is defined as
\begin{eqnarray}
\Omega(T, V, \mu)=-T \ln \mathcal{Z} .
\end{eqnarray}
Under the mean-field approximation, thermodynamic potential per unit volume, $\frac{\Omega}{V}$ in the chiral SU(3) model is given by the equation,
\begin{align}
    \frac{\Omega}{V}&=-\frac{\gamma_i T}{(2 \pi)^3} \sum_i \int d^3 k\left\{\ln \left(1+e^{-\beta\left[E_i^*(k)-\mu_i^*\right]}\right)+\ln \left(1+e^{-\beta\left[E_i^*(k)+\mu_i^*\right]}\right)\right\}\nonumber\\
&-\mathcal{L}_{v e c}-\mathcal{L}_0-\mathcal{L}_{S B} -V_{vac} .
\label{Eq_themo_pot1}
\end{align}
In the above expression, \(\gamma_i\) is the spin degeneracy factor having value $2$ for spin$-\frac{1}{2}$ and $4$ for spin$-\frac{3}{2}$ baryons, respectively. The effective single-particle energy is represented by \(E_i^*(k) = \sqrt{k^2 + m_i^{*2}}\). Furthermore, the effective chemical potential,  \(\mu_i^* = \mu_i - g_{\omega i} \omega - g_{\phi i} \phi - g_{\rho i} I_{3i} \rho\), where \(g_{\omega i}\), \(g_{\phi i}\), and \(g_{\rho i}\) are the coupling constants of the vector fields \(\omega\), \(\phi\), and \(\rho\) with the baryons, respectively. The chemical potential $\mu_i$ can be expressed in terms of baryon, isospin, and strangeness  chemical potentials, $\mu_B$, $\mu_I$ and $\mu_S$, though relation
\begin{equation}
	\mu_i = B_i \mu_B + I_{3i} \mu_I + S_i \mu_S .
	\label{che_pot}
\end{equation}
In the above, $B_i$  and $S_i$ represent the baryon and strangeness quantum numbers.  
In Eq.~(\ref{Eq_themo_pot1}), the term $V_{vac}$ is subtracted to make the vacuum energy zero.

The coupled equations of motion for scalar fields $\sigma, \zeta, \delta$, the dilaton field, $\chi$, and the vector fields $\omega$, $\rho$ and $\phi$  are obtained by minimizing the total thermodynamics potential and are expressed as
\begin{align}
\label{Eq_sigma_eq1}
&k_0\chi^2\sigma-4 k_1\left(\sigma^2+\zeta^2+\delta^2\right) \sigma-2 k_2\left(\sigma^3+3 \sigma \delta^2\right) 
 -2 k_3 \chi \sigma \zeta-\frac{d}{3} \chi^4\left(\frac{2 \sigma}{\sigma^2-\delta^2}\right)  \nonumber\\ &+\frac{\chi^2}{\chi_0^2} m_\pi^2 f_\pi 
 =\sum_{i} g_{\sigma i} \rho_i^s ,
\end{align}
\begin{align}
& k_{0}\chi^{2}\zeta-4k_{1}\left( \sigma^{2}+\zeta^{2}+\delta^{2}\right)
\zeta-4k_{2}\zeta^{3}-k_{3}\chi\left( \sigma^{2}-\delta^{2}\right) 
-\frac{d}{3}\frac{\chi^{4}}{\zeta} \nonumber\\
&+\left(\frac{\chi}{\chi_{0}} \right)
^{2}\left[ \sqrt{2}m_{K}^{2}f_{K}-\frac{1}{\sqrt{2}} m_{\pi}^{2}f_{\pi}\right]
 =\sum_{i} g_{\zeta i}\rho_{i}^{s} ,
 \label{Eq_zeta11}
\end{align}
\begin{align}
\label{Eq_delta}
	&k_0 \chi^2 \delta-4 k_1\left(\sigma^2+\zeta^2+\delta^2\right) \delta-2 k_2\left(\delta^3+3 \delta \sigma^2\right)
 -2 k_3 \chi \delta \zeta-\frac{d}{3} \chi^4\left(\frac{2 \delta}{\sigma^2-\delta^2}\right)   
	=\sum_{i} g_{\delta i}I_{3i} \rho_i^s ,
\end{align}
\begin{align}
\label{Eq_chi}
&k_0 \chi\left(\sigma^2+\right.\left.\zeta^2+\delta^2\right)-k_3 \chi\left(\sigma^2-\delta^2\right) \zeta+\left[4 k_4+1-\ln \frac{\chi^4}{\chi_0^4}-\frac{4 d}{3} \ln \left(\frac{\left(\sigma^2-\delta^2\right) \zeta}{\sigma_0^2 \zeta_0}\right)\right]\chi^3\nonumber\\ &+ \frac{2 \chi}{\chi_0^2}\left[m_\pi^2 f_\pi \sigma+\left(\sqrt{2} m_K^2 f_K-\frac{1}{\sqrt{2}} m_\pi^2 f_\pi\right) \zeta\right]-\frac{\chi}{\chi_0^2}\left(m_\omega^2 \omega^2+m_\rho^2 \rho^2+m_\phi^2 \phi^2\right)=0 ,
\end{align}
\begin{align}
\label{Eq_omega_field}
	\frac{\chi^2}{\chi_0^2}\left(m_\omega^2 \omega\right)+4 g_4 \omega^3+12 g_4 \omega \rho^2=\sum_{i} g_{\omega i} \rho_i^v ,
\end{align}
\begin{align}
\label{Eq_rho_field}
	\frac{\chi^2}{\chi_0^2}\left(m_\rho^2 \rho\right)+4 g_4 \rho^3+12 g_4 \omega^2 \rho=\sum_{i} g_{\rho i}I_{3i} \rho_i^v  ,
\end{align}
\begin{align}
\label{Eq_phi_field}
	\frac{\chi^2}{\chi_0^2}\left(m_\phi^2 \phi\right)+8 g_4 \omega^3=\sum_{i} g_{\phi_i} \rho_i^v .
\end{align}

In the above equations, the model parameters $k_{0}$, $k_{2}$ and $k_{4}$ are fitted to reproduce the vacuum values of $\sigma$, $\zeta$, $\chi$ fields and the remaining constant $k_{1}$ is fixed to produce the effective nucleon mass at saturation density around 0.65$m_N$  and $k_{3}$ is the constrained by $\eta$ and $\eta '$ masses \cite{Papazoglou:1998vr}.
 %The parameters $\chi_0$ and \(g_4\) are adjusted in the model to obtain appropriate   nuclear saturation properties for symmetric nuclear matter \((E/{\rho_B} - m_N = -16 \, \text{MeV}, \, E_{\text{sym}} = 32 \, \text{MeV},   \, \text{at} \, \rho_0 = 0.15 \, \text{fm}^{-3})\).
%, we have adjusted \(\chi_0\) and \(g_4\). 
Also, $\rho_i^v$ and $\rho_i^s$ represent the vector and scalar densities of the $i^{th}$ the baryons, respectively, and are written as  
\begin{eqnarray}
\rho_{i}^{v} = \gamma_{i}\int\frac{d^{3}k}{(2\pi)^{3}}  
\Bigg(\frac{1}{1+\exp\left[\beta(E^{\ast}_i(k) 
-\mu^{*}_{i}) \right]}-\frac{1}{1+\exp\left[\beta(E^{\ast}_i(k)
+\mu^{*}_{i}) \right]}\Bigg) ,
\label{rhov0}
\end{eqnarray}

 and

\begin{eqnarray}
\rho_{i}^{s} = \gamma_{i}\int\frac{d^{3}k}{(2\pi)^{3}} 
\frac{m_{i}^{*}}{E^{\ast}_i(k)} \Bigg(\frac{1}{1+\exp\left[\beta(E^{\ast}_i(k) 
-\mu^{*}_{i}) \right]}+\frac{1}{1+\exp\left[\beta(E^{\ast}_i(k)
+\mu^{*}_{i}) \right]}\Bigg) .
\label{rhos0}
\end{eqnarray}
 The system of non-linear equations, including Eqs. (\ref{Eq_sigma_eq1})–(\ref{Eq_phi_field}) and Eq. \ref{che_pot}, are solved for the medium having octet and decuplet baryons, at various values of baryon density ($\rho_{B}$), temperature ($T$), isospin asymmetry ($I_a$), and strangeness fraction ($f_{s}$) of the medium. The parameters $I_a$ and $f_{s}$ are defined as $-\frac{\sum_i I_{3i} \rho^{v}_{i}}{\rho_{B}}$ and $\frac{\sum_i \vert S_{i} \vert \rho^{v}_{i}}{\rho_{B}}$, respectively. %Here,  $s_{i}$ represents the number of strange quarks in the $i^{\text{th}}$ baryon.
\subsection{$K$ and $\bar{K}$ interactions in the chiral SU(3) model}
\label{subsec2.2}
In this section, we study the interactions of kaons, $K$ $\left(K^+, K^0\right)$, and antikaons, $\bar K \left(K^-, \bar{K^0} \right)$, in the isospin asymmetric dense resonance medium. 
The interactions of $K$ and $\bar{K}$ mesons with nucleons, hyperons, resonance baryons, and scalar meson fields ($\sigma$, $\zeta$, $\delta$) within the chiral SU(3) framework give rise to their medium-modified energies.
 The interaction Lagrangian density is  written as
\begin{equation}
\label{Eq_kBD_gen1}
\mathcal L _{KB} = \mathcal{L}_{OW}+\mathcal{L}_{DW}+ \mathcal L _{mass} + {\mathcal L}_{{\mathrm{1st\, range\, term}}} + {\mathcal L }_{d_1}^{BM} + {\mathcal L }_{d_2}^{BM} , 
\end{equation}
% which includes the baryon-meson interaction, the scalar meson exchange, and range terms.
where $\mathcal{L}_{OW}$ and $\mathcal{L}_{DW}$ are Lagrangian densities having interaction terms of $K$ and $\bar{K}$ mesons with  octet and decuplet baryons,  respectively.
The Lagrangian density, $\mathcal{L}_{OW}$, is the kinetic term of octet baryons (first term of Eq.~(\ref{Eq_kinetic_2})). As described earlier, in the kinetic term of baryons, the pseudoscalar $K$ and $\bar{K}$ mesons appear through the unitary transformation operator $u$ defined in terms of pseudoscalar meson octet matrix.
Explicitly, for $K$ and $\bar{K}$ mesons, the expression of $\cal L _{OW}$ (Weinberg Tomozawa term) is given by \cite{Mishra:2008dj}
\begin{eqnarray}
\cal L _{OW} & = & -\frac {i}{4 f_K^2} \Big [\Big ( 2 \bar p \gamma^\mu p
+\bar n \gamma ^\mu n -\bar {\Sigma^-}\gamma ^\mu \Sigma ^-
+\bar {\Sigma^+}\gamma ^\mu \Sigma ^+
- 2\bar {\Xi^-}\gamma ^\mu \Xi ^-
- \bar {\Xi^0}\gamma ^\mu \Xi^0 \Big)
\nonumber \\
& \times &
\Big(K^- (\partial_\mu K^+) - (\partial_\mu {K^-})  K^+ \Big )
\nonumber \\
& + &
\Big ( \bar p \gamma^\mu p
+ 2\bar n \gamma ^\mu n +\bar {\Sigma^-}\gamma ^\mu \Sigma ^-
-\bar {\Sigma^+}\gamma ^\mu \Sigma ^+
- \bar {\Xi^-}\gamma ^\mu \Xi ^-
- 2 \bar {\Xi^0}\gamma ^\mu \Xi^0 \Big)
\nonumber \\
& \times &
\Big(\bar {K^0} (\partial_\mu K^0) - (\partial_\mu {\bar {K^0}})  K^0 \Big )
\Big ].
\end{eqnarray}
To obtain the Lagrangian density, $\mathcal{L}_{DW}$, describing the
interactions of $K$ and $\bar{K}$  mesons with all the decuplet baryons
($\Delta^{++}$, $\Delta^{\pm,0}$, $\Sigma^{*\pm,0}$, $\Xi^{*-,0}$, $\Omega^{-}$) defined through baryon field $T_{a b c}^\mu$,
we consider the expression
\cite{sarkar2005baryonic,Geng:2009hh,Holmberg:2018dtv}  
  \begin{equation}
\label{Eq_Decu_Tb1}
\mathcal{L}_{DW}=-i \bar{T}^\mu \mathcal{\cancel{D}} {T}_\mu .
\end{equation}
In the above equation, ${D}^\nu$ is the covariant derivative operating over decuplet baryon field through relation
\begin{eqnarray}
\mathcal{D}^\nu T_{a b c}^\mu=\partial^\nu T_{a b c}^\mu+i\left(\Gamma^\nu\right)_a^d T_{d b c}^\mu+i\left(\Gamma^\nu\right)_b^d T_{a d c}^\mu+i\left(\Gamma^\nu\right)_c^d T_{a b d}^\mu ,
\end{eqnarray}
where $\mu$ is the Lorentz index and $a, b, c, d$ are the $S U(3)$ indices.
% The vector current $\Gamma^\nu$ is specified as
%\begin{equation}
%\label{gammamu}
%\Gamma_{\nu} =-\frac{i}{2} \left[u^{\dagger}(\partial_{\nu}u) - (\partial_{\nu} u^\dagger) u 
%+u (\partial_{\nu} u^\dagger) - (\partial_{\nu}u) u^{\dagger}  \right]. 
%\end{equation}
% \begin{eqnarray}
% \Gamma^\nu=\frac{1}{2}\left(\xi \partial^\nu \xi^{\dagger}+\xi^{\dagger} \partial^\nu \xi\right)
% \end{eqnarray}
% with
% \begin{eqnarray}
% \xi^2=U=e^{i \sqrt{2} \Phi / f}
% \end{eqnarray}
% where $\Phi$ is the ordinary $3 \times 3$ matrix of fields for the pseudoscalar mesons  and $f=93$ $\mathrm{MeV}$. 
% In terms of matrix form, the interaction Lagrangian for decuplet-meson interaction can  be provided as
% \begin{equation}
% \label{eq29}
% (\bar{T} \cdot T)_{a d}=\sum_{b, c} \bar{T}^{a b c} T_{d b c}
% \end{equation}
% as
% \begin{eqnarray}
% \mathcal{L}=3 i {Tr}\left\{\bar{T} \cdot T \Gamma^{0 T}\right\}
% \end{eqnarray}
% where $\Gamma^{0 T}$ is the transposed matrix of $\Gamma^0$, with $\Gamma^\nu$ given, up to two meson fields, by
% \begin{eqnarray}
% \Gamma^\nu=\frac{1}{4 f^2}\left(\Phi \partial^\nu \Phi-\partial^\nu \Phi \Phi\right) .
% \end{eqnarray}
To obtain the explicit expression for the interactions of decuplet members with $K$ and $\bar{K}$ mesons, following identification of the $S U(3)$ component of decuplet baryon field, $T$, to the physical states  is made \cite{sarkar2005baryonic,Geng:2009hh,Holmberg:2018dtv}
$$
	 T^{111}=\Delta^{++}, T^{112}=\frac{1}{\sqrt{3}} \Delta^{+}, T^{122}=\frac{1}{\sqrt{3}} \Delta^0, T^{222}=\Delta^{-}, T^{113}=\frac{1}{\sqrt{3}} \Sigma^{*+}, T^{123}=\frac{1}{\sqrt{6}} \Sigma^{* 0}, \nonumber \\
$$
$$
  T^{223}=\frac{1}{\sqrt{3}} \Sigma^{*-}, T^{133}=\frac{1}{\sqrt{3}} \Xi^{* 0}, T^{233}=\frac{1}{\sqrt{3}} \Xi^{*-}, T^{333}=\Omega^{-} .
$$
\newline
Using the above, we obtain the following expression from Eq.~(\ref{Eq_Decu_Tb1}) 
%\begin{eqnarray}
%\cal L _{OW} & = & -\frac {i}{4 f_K^2} \Big [\Big ( 2 \bar p \gamma^\mu p
%+\bar n \gamma ^\mu n -\bar {\Sigma^-}\gamma ^\mu \Sigma ^-
%+\bar {\Sigma^+}\gamma ^\mu \Sigma ^+
%- 2\bar {\Xi^-}\gamma ^\mu \Xi ^-
%- \bar {\Xi^0}\gamma ^\mu \Xi^0 \Big)
%\nonumber \\
%& \times &
%\Big(K^- (\partial_\mu K^+) - (\partial_\mu {K^-})  K^+ \Big )
%\nonumber \\
%& + &
%\Big ( \bar p \gamma^\mu p
%+ 2\bar n \gamma ^\mu n +\bar {\Sigma^-}\gamma ^\mu \Sigma ^-
%-\bar {\Sigma^+}\gamma ^\mu \Sigma ^+
%- \bar {\Xi^-}\gamma ^\mu \Xi ^-
%- 2 \bar {\Xi^0}\gamma ^\mu \Xi^0 \Big)
%\nonumber \\
%& \times &
%\Big(\bar {K^0} (\partial_\mu K^0) - (\partial_\mu {\bar {K^0}})  K^0 \Big )
%\Big ]
%\end{eqnarray}
 \begin{align}
\cal L _{DW} & =  -\frac {3 i}{4 f_K^2} \Big [\Big (3 \bar {\Delta^{++}} \gamma^\mu \Delta^{++} +2 \bar {\Delta^{+}} \gamma^\mu \Delta^{+} + \bar {\Delta^{0}} \gamma^\mu \Delta^{0}- \bar {\Sigma^{*-}}\gamma ^\mu \Sigma ^{*-}+ \bar {\Sigma^{*+}}\gamma
^\mu \Sigma ^{*+}
\nonumber\\
& -2 \bar {\Xi^{*-}}\gamma ^\mu \Xi ^{*-} - \bar {\Xi^{*0}}\gamma ^\mu \Xi ^{*0} - 3\bar {\Omega^-}\gamma ^\mu \Omega ^-  \Big)
\nonumber \\
& \times 
\Big(K^- (\partial_\mu K^+) - (\partial_\mu {K^-})  K^+ \Big )
\nonumber \\
& + 
\Big ( 3 \bar {\Delta^{-}} \gamma^\mu \Delta^{-} + \bar {\Delta^{+}} \gamma^\mu \Delta^{+} +2 \bar {\Delta^{0}} \gamma^\mu \Delta^{0} - \bar {\Sigma^{*+}}\gamma ^\mu \Sigma ^{*+} + \bar {\Sigma^{*-}}\gamma ^\mu \Sigma ^{*-}
\nonumber\\
& -  \bar {\Xi^{*-}}\gamma ^\mu \Xi ^{*-} -2 \bar {\Xi^{*0}}\gamma ^\mu \Xi ^{*0} -3\bar {\Omega^-}\gamma ^\mu \Omega ^- \Big)
\nonumber \\
& \times 
\Big(\bar {K^0} (\partial_\mu K^0) - (\partial_\mu {\bar {K^0}})  K^0 \Big )
\Big ].
\end{align}
%for strange hadronic matter \cite{mishra2009kaon, kumar2020phi} and resonance matter respectively.
 The third term in Eq.~(\ref{Eq_kBD_gen1}) is the scalar meson exchange term (mass term), which is obtained from the Lagrangian density given in Eq.~(\ref{Eq_SB_chiral1}), when the second term in the expansion of operator $u$ is considered. This gives the following interaction of $K$ and $\bar{K}$ mesons with scalar fields $\sigma, \zeta$ and $\delta$ 
\begin{equation}
\label{33}
 {\mathcal L}_{mass}  = \frac{m_K^2}{2f_K} \left[\left(\sigma + \sqrt{2} \zeta + \delta\right) K^+ K^- +
 \left(\sigma + \sqrt{2} \zeta - \delta\right) K^0 \bar{K^0} \right].
\end{equation}
%where $X$ is the scalar meson multiplet and $A_{p}$ given as,
%\begin{eqnarray}
%A_p&=&1/\sqrt{2} {\mathrm{diag}} (m_{\pi}^2 f_{\pi},
% m_\pi^2 f_\pi,
%  2 m_K^2 f_K 
%-m_{\pi}^2 f_\pi) .
%\end{eqnarray}
In Eq.~(\ref{Eq_kBD_gen1}), the  fourth term is the kinetic energy term for the pseudoscalar $K$ and $\bar{K}$ mesons (general expression is third term in Eq.~(\ref{Eq_kinetic_2})) and is given as,
\begin{equation}
\label{pikin}
 {\mathcal L}_{{\mathrm{1st\, range\, term}}} = - \frac{1}{f_K} \left[\left(\sigma + \sqrt{2} \zeta + \delta\right) (\partial_\mu K^+) (\partial^\mu K^-) +
  \left(\sigma + \sqrt{2} \zeta - \delta\right) (\partial_\mu K^0) (\partial^\mu \bar{K^0}) \right]. 
\end{equation}
%where, $u_{\mu}$ is given as,
%\begin{equation}
%u_{\mu} =-\frac{i}{2} \left[u^{\dagger}(\partial_{\mu}u) 
%-u (\partial_{\mu}u^\dagger) \right]. 
%\end{equation}
In the chiral SU(3) model, the range terms $d_1$ and $d_2$ were introduced to incorporate the meson- octet baryon interactions at next-to-leading order \cite{Mishra:2004te, Mishra:2006wy}.
The next-to-leading order terms are also defined for the interactions of decuplet baryons with mesons in Ref.\cite{Geng:2009hh,Holmberg:2018dtv}. In the present work we take into account such interaction Lagrangian densities for the octet as well as decuplet baryons through the
last two terms in Eq.~(\ref{Eq_kBD_gen1}) whose general  expressions are written as
\begin{equation}
\label{Eq20}
{\cal L }_{d_1}^{BM} =\frac {d_1}{2} Tr (u_\mu u ^\mu) \Big[Tr( \bar B B) + Tr( \bar T T) \Big ] ,
\end{equation}
and
\begin{equation}
\label{Eq21}
{\cal L }_{d_2}^{BM} =d_2 \Big [Tr (\bar B u_\mu u ^\mu B) + Tr (\bar T u_\mu u ^\mu T) \Big ].
\end{equation}
From above we obtain following 
\begin{eqnarray}
{\cal L }_{d_1}^{BM} & = & \frac {d_1}{2 f_K^2}(\bar p p +\bar n n +\bar {\Lambda^0}{\Lambda^0}
+\bar {\Sigma ^+}{\Sigma ^+}
+\bar {\Sigma ^0}{\Sigma ^0}
+\bar {\Sigma ^-}{\Sigma ^-}
+\bar {\Xi ^-}{\Xi ^-}
+\bar {\Xi ^0}{\Xi ^0}
+ 3 (\bar {\Delta^{++}}\Delta^{++}
\nonumber\\
& + & \bar {\Delta^{+}} \Delta^{+} + \bar {\Delta^{-}} \Delta^{-} + \bar {\Delta^{0}} \Delta^{0} + \bar {\Sigma ^{*+}}{\Sigma ^{*+}}
+\bar {\Sigma ^{*0}}{\Sigma ^{*0}}
+\bar {\Sigma ^{*-}}{\Sigma ^{*-}}
+\bar {\Xi ^{*-}}{\Xi ^{*-}}
+\bar {\Xi ^{*0}}{\Xi ^{*0}} 
+\bar {\Omega ^{-}}{\Omega ^{-}})
 )\nonumber \\
&\times & \big ( (\partial _\mu {K^+})(\partial ^\mu {K^-})
+(\partial _\mu {K^0})(\partial ^\mu {\bar {K^0}})
\big )
\end{eqnarray}
and
\begin{eqnarray}
{\cal L }_{d_2}^{BM}&=& \frac {d_2}{2 f_K^2} \Big [
(\bar p p+\frac {5}{6} \bar {\Lambda^0}{\Lambda^0}
+\frac {1}{2} \bar {\Sigma^0}{\Sigma^0}
+\bar {\Sigma^+}{\Sigma^+}
+\bar {\Xi^-}{\Xi^-}
+\bar {\Xi^0}{\Xi^0}
+3(\bar {\Delta^{++}} \Delta^{++} + \frac {2}{3}  \bar {\Delta^{+}} \Delta^{+}
\nonumber \\
& + & \frac {1}{3} \bar {\Delta^{0}} \Delta^{0} + \bar {\Sigma ^{*+}}{\Sigma ^{*+}}
+ \frac {2}{3} \bar {\Sigma ^{*0}}{\Sigma ^{*0}}
+ \frac {1}{3} \bar {\Sigma ^{*-}}{\Sigma ^{*-}}
+ \frac {2}{3} \bar {\Xi ^{*-}}{\Xi ^{*-}}
+\bar {\Xi ^{*0}}{\Xi ^{*0}} 
+\bar {\Omega ^{-}}{\Omega ^{-}}
)) (\partial_\mu K^+)(\partial^\mu K^-) 
\nonumber \\
 &+ &(\bar n n
+\frac {5}{6} \bar {\Lambda^0}{\Lambda^0}
+\frac {1}{2} \bar {\Sigma^0}{\Sigma^0}
+\bar {\Sigma^-}{\Sigma^-}
+\bar {\Xi^-}{\Xi^-}
+\bar {\Xi^0}{\Xi^0}
+ 3(\frac {1}{3}  \bar {\Delta^{+}} \Delta^{+} + \frac {2}{3}  \bar {\Delta^{0}} \Delta^{0} +\bar {\Delta^{-}} \Delta^{-}
\nonumber \\
&+& \frac {1}{3}  \bar {\Sigma ^{*+}}{\Sigma ^{*+}}
+ \frac {2}{3} \bar {\Sigma ^{*0}}{\Sigma ^{*0}}
+ \bar {\Sigma ^{*-}}{\Sigma ^{*-}}
+\frac {2}{3} \bar {\Xi ^{*0}}{\Xi ^{*0}}
+ \bar {\Xi ^{*-}}{\Xi ^{*-}}
+{\bar {\Omega ^{-}}}{\Omega ^{-}}
)) (\partial_\mu K^0)(\partial^\mu {\bar {K^0}})
\Big ].
\end{eqnarray}
%The interaction Lagrangian density for $K$ and $\bar K$ with baryons is given by 

The dispersion relation to compute the in-medium masses of $K$ ($K^+$, $K^0$) and $\bar K$ ($K^-$, $\bar {K^0}$) mesons is obtained from Eq.~(\ref{Eq_kBD_gen1}) using Euler-Lagrange equation of motion and applying the Fourier transformation. Thus,  we have
\begin{equation}
-\omega^2+ {\vec k}^2 + m_{K (\bar K)}^2 -\Pi^*(\omega, |\vec k|)=0,
\label{eq_self_energy}
\end{equation}
where $\Pi^*(\omega, |\vec k|)$ represents the in-medium self-energy for $K$ and $\bar K$ mesons. The self-energy for kaon isospin doublet $K$ ($K^+$, $K^0$), is given by
\begin{eqnarray}
\Pi^*_K (\omega, |\vec k|) &= & -\frac {1}{4 f_K^2}\Big [3 (\rho_p +\rho_n)
\pm (\rho_p -\rho_n) \pm 2 (\rho_{\Sigma^+}-\rho_{\Sigma^-})
-\big ( 3 (\rho_{\Xi^-} +\rho_{\Xi^0}) \pm (\rho_{\Xi^-} 
\nonumber\\
&-&\rho_{\Xi^0})\big) +
3\Big\{\big(3(\rho_{\Delta^+} + \rho_{\Delta^0})\pm (\rho_{\Delta^+} - \rho_{\Delta^0}) \big) \pm 2 (\rho_{\Sigma^{*+}}- \rho_{\Sigma^{*-}}) \nonumber\\
&-& \big(3(\rho_{\Xi^{*-}} + \rho_{\Xi^{*0}})\pm (\rho_{\Xi^{*-}} - \rho_{\Xi^{*0}}) \big)
- 6 \rho_{\Omega^-} +6 \big( a \rho_{\Delta^{++}} + b \rho_{\Delta^-})   \big)\Big\} \Big ] \omega \nonumber\\
&+& \frac {m_K^2}{2 f_K} (\sigma ' +\sqrt 2 \zeta ' \pm \delta ')
\nonumber \\ & +& \Big [- \frac {1}{f_K}
(\sigma ' +\sqrt 2 \zeta ' \pm \delta ')
+\frac {d_1}{2 f_K ^2} \Big ({\rho^s}_{p} +{\rho^s}_{n}
+{\rho^s} _{\Lambda^0}+ {\rho^s} _{\Sigma^+}+{\rho^s} _{\Sigma^0}
+{\rho^s} _{\Sigma^-} \nonumber\\
&+&{\rho^s} _{\Xi^-} +{\rho^s} _{\Xi^0}+ 3\Big \{{\rho^s}_{\Delta^{++}} + {\rho^s}_{\Delta^+}+ {\rho^s}_{\Delta^0} + {\rho^s}_{\Delta^-} + {\rho^s} _{\Sigma^{*+}}+{\rho^s} _{\Sigma^{*0}} + {\rho^s} _{\Sigma^{*-}} \nonumber\\
&+& {\rho^s} _{\Xi^{*-}} +{\rho^s} _{\Xi^{*0}} + {\rho^s}_{\Omega^-} \Big  \}\Big)
+\frac {d_2}{4 f_K ^2} \Big (({\rho^s} _p +{\rho^s} _n)
\pm   ({\rho^s} _p -{\rho^s} _n)
+{\rho^s} _{\Sigma ^0}+\frac {5}{3} {\rho^s} _{\Lambda^0} \nonumber\\
&+& ({\rho^s} _{\Sigma ^+} + {\rho^s} _{\Sigma ^-})
\pm ({\rho^s} _{\Sigma ^+}-{\rho^s} _{\Sigma ^-})
 +  2 {\rho^s} _ {\Xi^-} + 2 {\rho^s} _ {\Xi^0} + 3 \Big\{ 2 ( a \rho^s_{\Delta^{++}} + b \rho^s_{\Delta^-}) \nonumber\\ 
 &+& \frac{1}{3} (3(\rho^s_{\Delta^+} + \rho^s_{\Delta^0})\pm (\rho^s_{\Delta^+} - \rho^s_{\Delta^0})) + \frac{4}{3} \rho^s_{\Sigma^{*0}} + \frac{1}{3} (4(\rho^s_{\Sigma^{*+}} 
 + \rho^s_{\Sigma^{*-}})\pm 2 (\rho^s_{\Sigma^{*+}} - \rho^s_{\Sigma^{*-}})) \nonumber\\
 &+& \frac{1}{3} (5(\rho^s_{\Xi^{*0}} + \rho^s_{\Xi^{*-}})\pm (\rho^s_{\Xi^{*0}} - \rho^s_{\Xi^{*-}})) +2 \rho^s_{\Omega^-}\Big\}
\Big )
\Big ]
(\omega ^2 - {\vec k}^2),
\label{sek}
\end{eqnarray}
where $\sigma '$, $\zeta '$ and $\delta '$ indicate the deviation of the fields, $\sigma$, $\zeta$, and $\delta$, from their vacuum expectation value, $\sigma_{0}$, $\zeta_{0}$, and $\delta_{0}$. 
For the antikaon isospin doublet $\bar{K}$ ($K^-$, $\bar {K^0}$), following expression is obtained. 
\begin{eqnarray}
\Pi^*_{\bar K} (\omega, |\vec k|) &= & \frac {1}{4 f_K^2}\Big [3 (\rho_p +\rho_n)
\pm (\rho_p -\rho_n) \pm 2 (\rho_{\Sigma^+}-\rho_{\Sigma^-})
-\big ( 3 (\rho_{\Xi^-} +\rho_{\Xi^0}) \pm (\rho_{\Xi^-} 
\nonumber\\
&-&\rho_{\Xi^0})\big) +
3\Big\{\big(3(\rho_{\Delta^+} + \rho_{\Delta^0})\pm (\rho_{\Delta^+} - \rho_{\Delta^0}) \big) \pm 2 (\rho_{\Sigma^{*+}}- \rho_{\Sigma^{*-}}) \nonumber\\
&-& \big(3(\rho_{\Xi^{*-}} + \rho_{\Xi^{*0}})\pm (\rho_{\Xi^{*-}} - \rho_{\Xi^{*0}}) \big)
- 6 \rho_{\Omega^-} +6 \big( a \rho_{\Delta^{++}} + b \rho_{\Delta^-})   \big)\Big\} \Big ] \omega \nonumber\\
&+& \frac {m_K^2}{2 f_K} (\sigma ' +\sqrt 2 \zeta ' \pm \delta ')
%\end{eqnarray*}
%\begin{eqnarray}
\nonumber \\ & +& \Big [- \frac {1}{f_K}
(\sigma ' +\sqrt 2 \zeta ' \pm \delta ')
+\frac {d_1}{2 f_K ^2} \Big ({\rho^s}_{p} +{\rho^s}_{n}
+{\rho^s} _{\Lambda^0}+ {\rho^s} _{\Sigma^+}+{\rho^s} _{\Sigma^0}
+{\rho^s} _{\Sigma^-} \nonumber\\
&+&{\rho^s} _{\Xi^-} +{\rho^s} _{\Xi^0}+ 3\Big \{{\rho^s}_{\Delta^{++}} + {\rho^s}_{\Delta^+}+ {\rho^s}_{\Delta^0} + {\rho^s}_{\Delta^-} + {\rho^s} _{\Sigma^{*+}}+{\rho^s} _{\Sigma^{*0}} + {\rho^s} _{\Sigma^{*-}} \nonumber\\
&+& {\rho^s} _{\Xi^{*-}} +{\rho^s} _{\Xi^{*0}} + {\rho^s}_{\Omega^-} \Big  \}\Big)
+\frac {d_2}{4 f_K ^2} \Big (({\rho^s} _p +{\rho^s} _n)
\pm   ({\rho^s} _p -{\rho^s} _n)
+{\rho^s} _{\Sigma ^0}+\frac {5}{3} {\rho^s} _{\Lambda^0} \nonumber\\
&+& ({\rho^s} _{\Sigma ^+} + {\rho^s} _{\Sigma ^-})
\pm ({\rho^s} _{\Sigma ^+}-{\rho^s} _{\Sigma ^-})
 +  2 {\rho^s} _ {\Xi^-} + 2 {\rho^s} _ {\Xi^0} + 3 \Big\{ 2 ( a \rho^s_{\Delta^{++}} + b \rho^s_{\Delta^-}) \nonumber\\ 
 &+& \frac{1}{3} (3(\rho^s_{\Delta^+} + \rho^s_{\Delta^0})\pm (\rho^s_{\Delta^+} - \rho^s_{\Delta^0})) + \frac{4}{3} \rho^s_{\Sigma^{*0}} + \frac{1}{3} (4(\rho^s_{\Sigma^{*+}} 
 + \rho^s_{\Sigma^{*-}})\pm 2 (\rho^s_{\Sigma^{*+}} - \rho^s_{\Sigma^{*-}})) \nonumber\\
 &+& \frac{1}{3} (5(\rho^s_{\Xi^{*0}} + \rho^s_{\Xi^{*-}})\pm (\rho^s_{\Xi^{*0}} - \rho^s_{\Xi^{*-}})) +2 \rho^s_{\Omega^-}\Big\}
\Big )
\Big ]
(\omega ^2 - {\vec k}^2).
\label{se_antik}
\end{eqnarray}
In the above equations, the ± symbol gives the self-energy for $K^+$ ($K^-$) and $K^0$ ($\bar K^0$) respectively. For the $K^{+}$ and $K^{-}$ , the parameters are set as $a=1$, $b=0$, whereas these are assigned as $a=0$, $b=1$ for $K^{0}$ and $\bar K^{0}$.

The in-medium masses of $K (\bar K)$ mesons in the dense resonance matter are calculated by Eq.~(\ref{eq_self_energy}) under the condition, $m_{K(\bar K)}^*=\omega(|\vec k|$=0) with the vacuum value of masses $K^+(K^-)$ and $K^0(\bar K^0)$ are taken to be 494 MeV and 498 MeV, respectively. The  parameters, $d_1$ and $d_2$, in the above equations, are taken as $ 2.56/m_K $ and $ 0.73/m_K $, respectively \cite{Mishra:2008kg,Mishra:2008dj}, fitted to empirical values  of kaon-nucleon  scattering length \cite{barnes1994kaon}.
The energies $\omega(k)$ of $K$ and $\bar K$ meson obtained from the dispersion relations are used to compute their optical potentials using expression %\cite{mishra2009kaon}. 
\begin{equation}
U(\omega, k) = \omega(k) - \sqrt{k^2 + m_K^2} .
\label{Eq_optK}
\end{equation} 

%Also, we explore how the optical potentials are influenced by the isospin asymmetry ($I_a$) and strangeness fraction ($f_{s}$) parameters within matter.

\section{Results and discussion} 
\label{results}
\begin{table}
\centering
\def\arraystretch{1.8}
\begin{tabular}{ccccccccc}
\hline \hline
 Baryon &${g_{\sigma i}}$& ${g_{\zeta i}}$  &   ${g_{\delta i}}$ &   ${g_{\omega i}}$ &${g_{\rho i}}$ &${g_{\phi i}}$\\ \hline
N&10.60 &  -0.47 &  2.49 &  13.33 &  5.49 &  0\\ 
$\Lambda$&5.31 & 5.80 & 0 & 8.89  & 0 & -6.29 \\ 
$\Sigma$&6.13 & 5.80 & 6.79 & 8.89 & 8.89 & -6.28\\ 
$\Xi$&3.68 & 9.14 & 2.36  & 4.44 & 4.44 & -12.56\\
$\Delta$&10.74 & 2.20 & 2.49 & 12.74 & 12.74 & 0 \\ 
$\Sigma^*$&8.20 & 5.80 & 3.47 & 8.50  & 8.50 & -6.01\\ 
$\Xi^*$&5.66 & 9.40 & 2.34 & 4.25 & 4.25 & -12.01\\ 
$\Omega$&3.12 & 15.19 & 0  & 0 & 0 & -18.02 \\
\hline \hline
\end{tabular}
%\captionsetup{justification=centering}
\caption{Coupling constants of scalar ($\sigma, \zeta$ and $\delta$) and vector
($\omega, \rho$ and $\phi$) fields with the octet and decuplet baryons.}
\label{coupling}
\end{table}

\begin{table}
\centering
\def\arraystretch{1.8}
\begin{tabular}{ccccccccc}
\hline \hline
 Baryon~ & $m_i$ (MeV) & $m^i_3$  &  $U_i$ (MeV) \\ \hline
N&939 & 0 & -77 \\ 
$\Lambda$&1115 & 0.5 & -20  \\ 
$\Sigma$&1193 & 1.88 & 31 \\ 
$\Xi$&1315 & 1.75 & -18 \\
$\Delta$&1232 & 0.68 & -80  \\ 
$\Sigma^*$&1385 & 1.4 & -64 \\ 
$\Xi^*$&1530 & 1.3 & -19 \\ 
$\Omega$&1672 & 39.25 & -25 \\
\hline \hline
\end{tabular}
%\captionsetup{justification=centering}
\caption{Values of vacuum masses, parameter $m_3^i$ and optical potentials (at $\rho_B = \rho_0$ and $\eta = f_s = 0$) of different baryons.}
\label{hyperon_pot}
\end{table}

\begin{table}
\centering
\begin{tabular}{ccccc}
\hline \hline
\vspace{0.5cm}
$k_0$=2.54 & $k_1$=1.35 & $k_2$=-4.78 & $k_3$=-2.08 & $k_4$=-0.29 \\ \vspace{0.5cm}
${f_\pi}=93.3$~MeV &${m_\pi}=139$~MeV & ${f_K}=122.14$MeV~ & ${m_K}=498$ ~MeV & ${m_N}=939.24$~ MeV \\ \vspace{0.5cm}
${m_\omega}=780.65$~ MeV & ${m_\rho}=761.06$~ MeV & ${m_\phi}=1019$ ~MeV &  $g_4=79.91$&$d=0.06$ \\ \vspace{0.5cm}
${\sigma_0}=-93.3$~MeV & ${\zeta_0}= -106.7$~MeV & ${\chi_0}=409.77$ ~MeV & 
 $\rho_0 = 0.15$~\text{fm}$^{-3}$ \\
\hline \hline
\end{tabular}
%\captionsetup{justification=centering}
\caption{Various parameters utilized in the present calculations.}
\label{constant parameters}
\end{table}

% \begin{table}[htp]
% \centering
% %\def\arraystretch{1.8}
% \begin{tabular}{ccccccc}
% \hline \hline

% $k_0$=2.53655899 & $k_1=1.35436$ & $k_2=-4.775199$ & $k_3=-20772557$ &$k_4=-0.21887 & ${f_\pi}=93.3$~MeV &
% ${f_k}=122.143$~MeV&\\ 
% ${m_\pi}=139$ MeV &${m_k}=498$ MeV & ${m_N}=939.243$ MeV ${ m_\omega}=780.65$ MeV &  ${ m_\rho}=761.06$ MeV  & ${ m_\phi}=1019.0$ MeV & $a=1$ \\
% % $\Sigma$&6.13 & 5.8 & 6.79 & 8.8843 & 8.8843 & -6.2822\\ 
% % $\Xi$&3.68 & 9.14 & 2.36  & 4.4422 & 4.4422 & -12.5643\\
% % $\Delta$&10.742 & 2.2 & 2.487 & 12.7428 & 12.7428 & 0 \\ 
% % $\Sigma^*$&8.2 & 5.8 & 3.4738 & 8.4952  & 8.4952 & -6.0070\\ 
% % $\Xi^*$&5.658 & 9.4 & 2.34 & 4.2476 & 4.2476 & -12.0140\\ 
% % $\Omega$&3.116 & 15.19 & 0  & 0 & 0 & -18.0210 \\
% \hline \hline
% \end{tabular}
% \caption{Different parameters used in the present calculations.}
% \label{constant parameters}
% \end{table}
The numerical results on in-medium masses and optical potential of $K$ and $\bar K$ mesons are presented in this section, derived from the medium-modified scalar fields $\sigma$, $\zeta$ and $\delta$ and vector fields $\omega$, $\rho$ and $\phi$ in isospin asymmetric dense resonance matter. In the following Sec.  \ref{IIIA},  we present the results for in-medium scalar fields corresponding to different values of isospin asymmetry and strangeness fractions. In Sec.  \ref{IIIB}, results of the effective masses and optical potential of $K$ and $\bar K$ mesons are presented. The coupling constants describing the strength of interactions of scalar and vector fields with octet and decuplet baryons  are given in Table \ref{coupling}. 
% listed the parameters utilized in this paper that differ from the coupling constants (previously elaborated in Sec.~\ref{method_chiral}).
%Before looking into the hadronic medium, it is essential to reproduce justifiable baryonic potentials within a hadronic matter. 
The couplings of vector fields with hyperons and decuplet baryons are fitted to the values of optical potentials of these baryons at nuclear saturation density $\rho_0$ in the symmetric nuclear matter.
The optical potential of a given baryon is defined  in terms of vector fields  through relation  
% \begin{equation}fiable
%      U_{i} = M^*_{i} - m_{i} + g_{\omega i} \omega + g_{\phi} \phi
% \end{equation}
\begin{equation}
 U_i= M^*_{i}- m_{i}+g_{\omega i} \omega+g_{\rho i} \rho+g_{\phi i} \phi.
 \end{equation}
 To achieve reasonable values for the baryon optical potential, an additional mass term in terms of parameter $m^i_3$ is incorporated into
 the effective mass $m^*_i$ of baryons  (defined in Eq.  \ref{mass_B}), resulting in \cite{Papazoglou:1998vr} \[
M^*_{i}= m^*_{i}+ m^i_3 \left( \sqrt{2} (\sigma_0 - \sigma) + (\zeta_0 - \zeta) \right).
\] 
The parameter $m^i_3$ is adjusted across all hyperons and resonance baryons to reproduce reasonable baryonic potentials $U_i$  at nuclear saturation density within symmetric nuclear matter. The values of vacuum masses and 
 optical potentials for different baryons are given  Table \ref{hyperon_pot}. The list of other parameters used in the present calculations is given in
 Table {\ref{constant parameters}}.

\subsection{In-medium scalar fields in resonance matter} 
\label{IIIA}
%The in-medium behavior of scalar fields ($\sigma$, $\zeta$, $\delta$, and $\chi$) in isospin asymmetric dense resonance matter is analyzed by plotting it with respect to the baryon density $\rho_B/\rho_0$ (in units of nuclear saturation density $\rho_0$) at finite temperatures in \textcolor{blue}{Figs.} \ref{fig1_sigmazeta} and \ref{fig5_deltachi}. 
% The results are shown for the isospin asymmetry ($\eta=0,0.3$) and strangeness fractions ($f_s=0,0.3,0.5$). 

\begin{figure}
    \centering
    \includegraphics[width=0.9\linewidth]{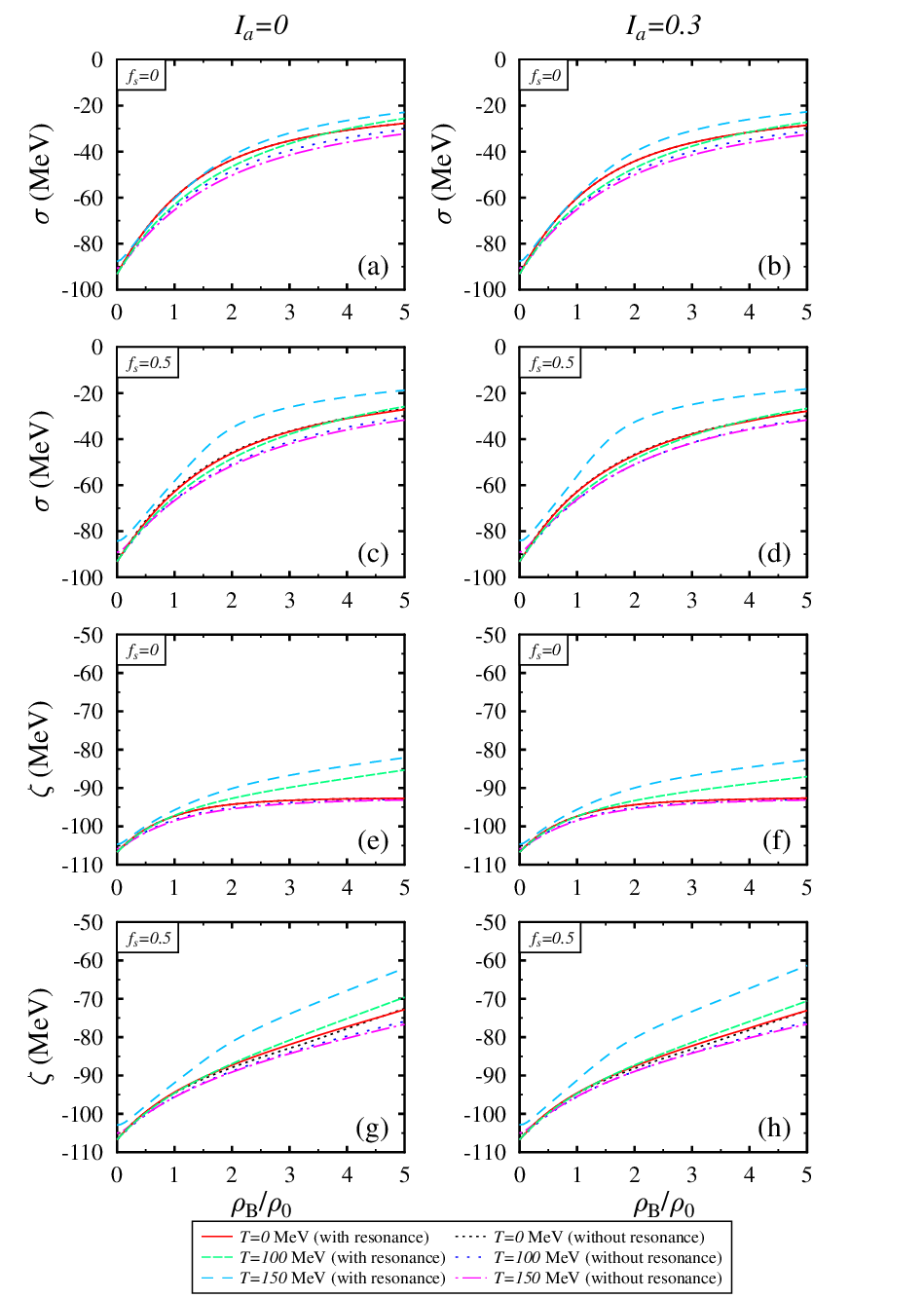}
    \caption{The scalar fields $\sigma$ and $\zeta$ are plotted as a function of the baryon density $\rho_B$ (in units of $\rho_0$), for isospin symmetric ($I_a=0$) and asymmetric ($I_a=0.3$) resonance matter.
The results are shown for strangeness fractions $f_s = 0$ (with nucleons and $\Delta$ baryons) [in the subplots (a), (b), (e) and (f)] and $f_s = 0.5$ (with nucleon, hyperons, and decuplet baryons) [in subplots (c), (d), (g) and (h)],
 at temperatures, $T = 0, 100$ and $150$ MeV. In each subplot, results are compared with the situation when resonance baryons are
 not considered in the medium.}
    \label{fig1_sigmazeta}
\end{figure}

In Fig., \ref{fig1_sigmazeta} we have shown the
variation of scalar fields $\sigma$ and
$\zeta$ as a function of baryon density ratio,
$\rho_B/\rho_0$, in resonance medium for isospin asymmetry $I_a = 0$ and $0.3$
and strangeness fractions $f_s = 0$ and $0.5$. In each subplot the results are 
plotted for temperatures $T = 0, 100$ and $150$ MeV and are compared with situation when resonance baryons are not included in the calculations.
%The impact of the isospin asymmetry ($I_a=0,0.3$) and strangeness fraction ($f_s=0,0.5$) is also observed in these %plots. 
In the subplots (a) and (b) of Fig. \ref{fig1_sigmazeta}, at $f_s=0$, we have shown the results for scalar field $\sigma$ as a function of $\rho_B/\rho_0$ for resonance matter, considering nucleons and $\Delta$ baryons.
 The magnitude of  non-strange scalar-isoscalar field $\sigma$ reduces as baryonic density $\rho_B$ increases for both symmetric ($I_a =0$) and asymmetric ($I_a =0.3$) resonance medium. In isospin symmetric medium, $I_a=0$, for baryon density below $4\rho_0$, the magnitude of scalar field
 $\sigma$ increases as the temperature of resonance matter is increased from $T=0$ to 100 MeV, whereas for baryon densities above $\sim 4\rho_0$, the magnitude of scalar field shifts to lower values as temperature of the medium is increased. Increasing the temperature from 100 to 150 MeV, shifts this
 baryon density value to lower side.
 For example, as shown in subplot (a), at  temperature $T=150$ MeV, the magnitude of scalar field $\sigma$ is lesser compared to $T = 0$ case after baryon density $\rho_B \sim 1.5\rho_0$. This is also possibly expected  due to the partial restoration of chiral symmetry at high temperature. 
When the $\Delta$ baryons are not considered in the medium, i.e., pure nuclear matter (shown as dotted lines in subplots (a) and (b)),  the magnitude
of scalar field $\sigma$ is larger at 
$T = 150 $ MeV also as compared to $T = 0$. The presence of $\Delta$ baryons
increases the attractive interactions
and leads to more decrease in the magnitude of fields at higher temperature.
 This behaviour of scalar field $\sigma$ also persists at finite isospin asymmetry of the medium (subplot (b) at $I_a=0.3$). For instance, at $\rho_B$ = $\rho_0$ (4$\rho_0$) and $f_s = 0$, for non-strange symmetric matter ($I_a = 0$), the values of $\sigma$ field change from its vacuum value $-93.30$ to $-60.0 (-30.6)$, $-63.0 (-30.0)$, and $-60.4 (-26.45)$ MeV, whereas for $I_a = 0.3$, these values change to $-60.2 (-31.54), -63.3 (-31.50)$, and $-59.54 (-26.0)$ MeV, at temperatures $T = 0, 100$, and $150$ MeV, respectively. 
 As one can observe, at low temperature increase of isospin asymmetry of medium causes less drop in the magnitude of scalar field $\sigma$.
\par
 In Figs. \ref{fig1_sigmazeta}(c) and (d),
 at $f_s=0.5$, results are shown considering all octet and decuplet baryons in the medium and significant modifications are found in the magnitude of the $\sigma$ field at the higher
 values of temperature. While shifting from strange symmetric to asymmetric matter, the $\sigma$ field exhibits the same pattern with more decrease in its magnitudes at high temperature.  For example, for $f_s=0.5$ at $\rho_B$ = $\rho_0 (4\rho_0)$ and $T=150$ MeV, the in-medium values of $\sigma$ field are found to be $-58.23 (-21.68)$ and $-56.07 (-20.95)$ MeV, for $I_a=0$ and $0.3$, respectively.
 Comparing the magnitude of these values with earlier quoted values of scalar 
 fields when only nucleons and $\Delta$ baryons were considered ($f_s =0$), we conclude that the consideration of all decuplet baryons further enhances the attractive interactions in the medium.
  By comparing these results with hyperonic matter (without resonance baryons), it is found that, in the hyperonic medium, the rate of drop in the magnitude of this field is reduced with an increase in the temperature
 of the medium. 
 
\begin{figure}
    \centering
    \includegraphics[width=0.6\linewidth]{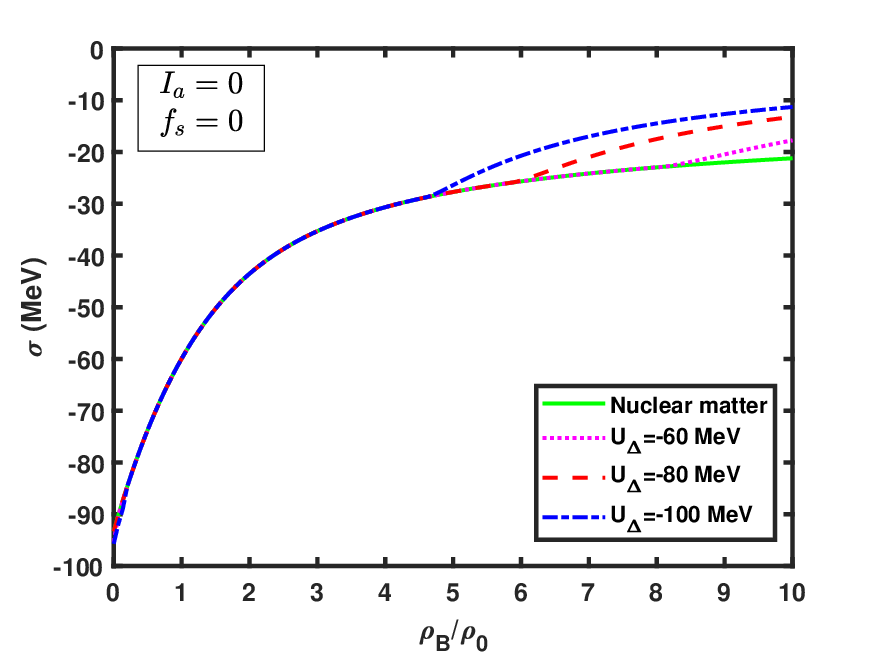}
    \captionsetup{justification=centering}
    \caption{The $\sigma$ field at various values of $\Delta$ baryon potentials for symmetric resonance matter.}
    \label{fig2_sigmaUD}
\end{figure}
\begin{figure}[hbt!]
\begin{subfigure}{.5\linewidth}
  (a)\includegraphics[width=0.9\linewidth]{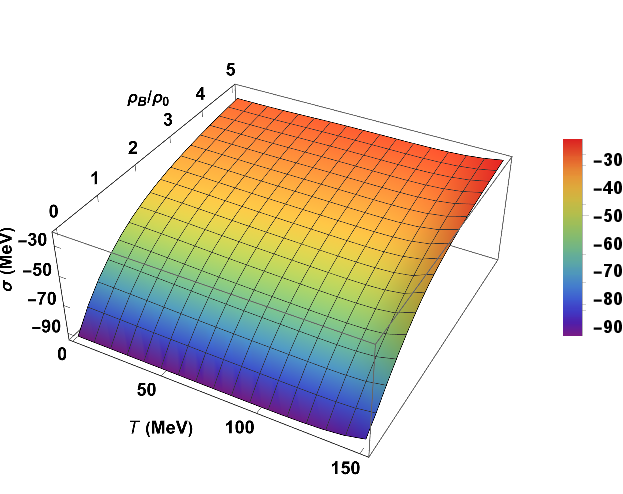}
 \end{subfigure}\hfill % <-- "\hfill"
\begin{subfigure}{.5\linewidth}
  (b)\includegraphics[width=0.9\linewidth]{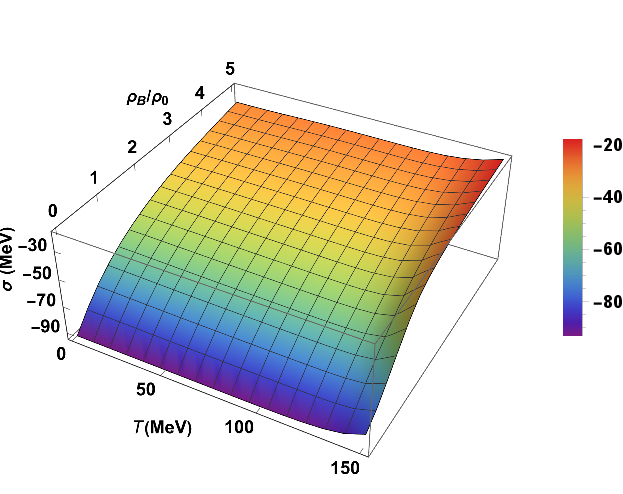}
 \end{subfigure}
\begin{subfigure}{.5\linewidth}
  (c)\includegraphics[width=0.9\linewidth]{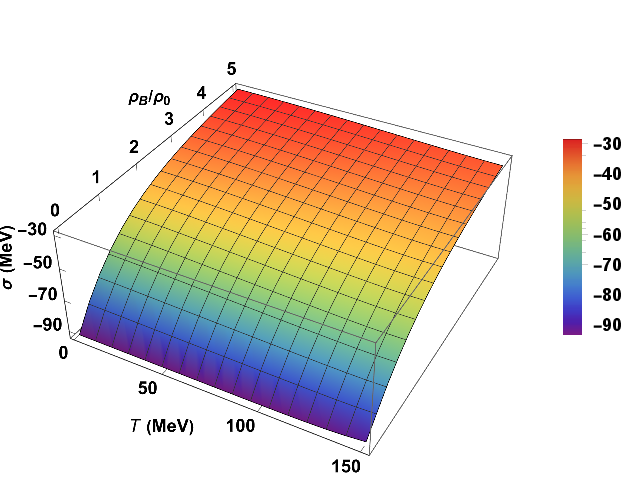}
  \end{subfigure}\hfill % <-- "\hfill"
\begin{subfigure}{.5\linewidth}
  (d)\includegraphics[width=0.9\linewidth]{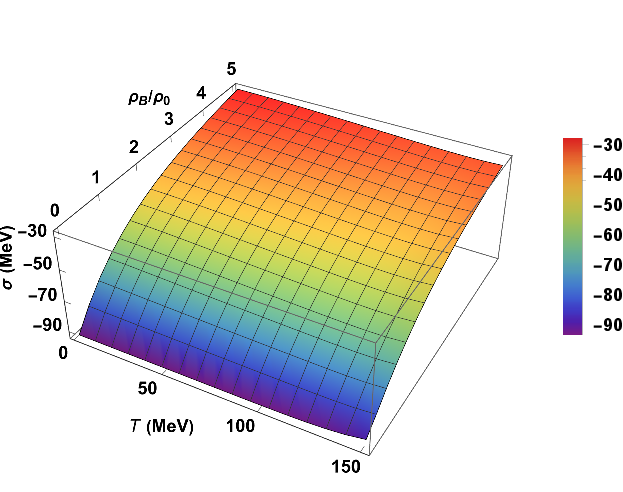}
  \end{subfigure}
\caption{The three-dimensional representation of scalar field $\sigma$ is shown as a function of baryon density ratio $\rho_B/\rho_0$ and temperature $T$ for isospin asymmetric medium ($I_a=0.3$). The subplots (a) and (c) correspond to $f_s = 0$, with (a) incorporating both nucleons and $\Delta$ baryons while (c) includes only nucleons. The subplots (b) and (d) are for $f_s=0.5$, where (b) considering all octet and decuplet baryons and (d) involves octet baryons only.}
\label{fig3_sigma3D}
\end{figure}

%%%%%%%%%%%%%%%%%%%
\begin{figure}[hbt!]
\begin{subfigure}{.5\linewidth}
  (a)\includegraphics[width=0.9\linewidth]{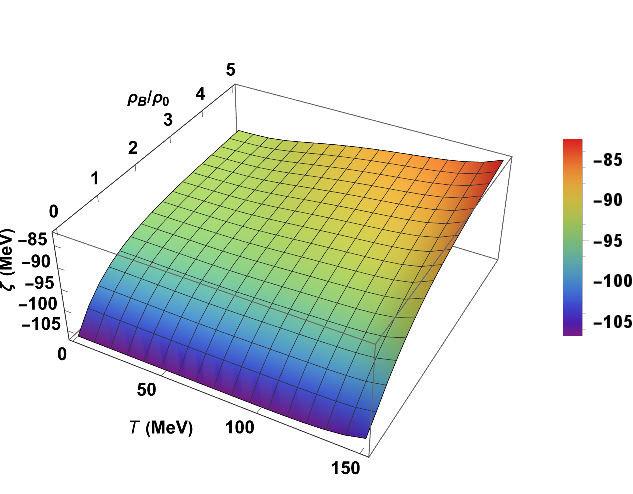}
 \end{subfigure}\hfill % <-- "\hfill"
\begin{subfigure}{.5\linewidth}
  (b)\includegraphics[width=0.9\linewidth]{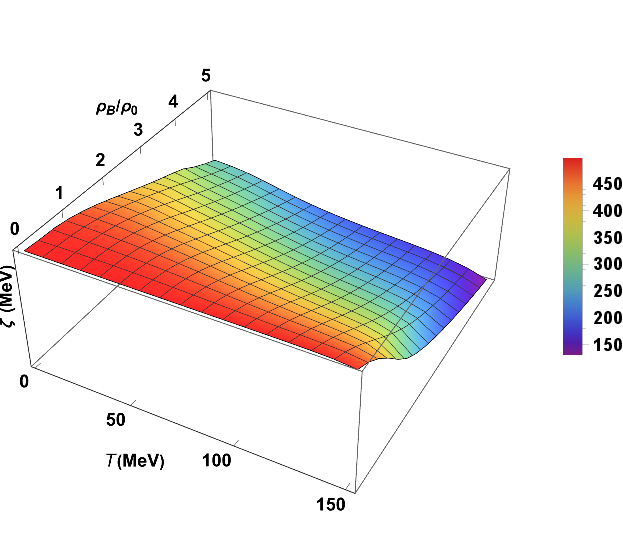}
 \end{subfigure}
\begin{subfigure}{.5\linewidth}
  (c)\includegraphics[width=0.9\linewidth]{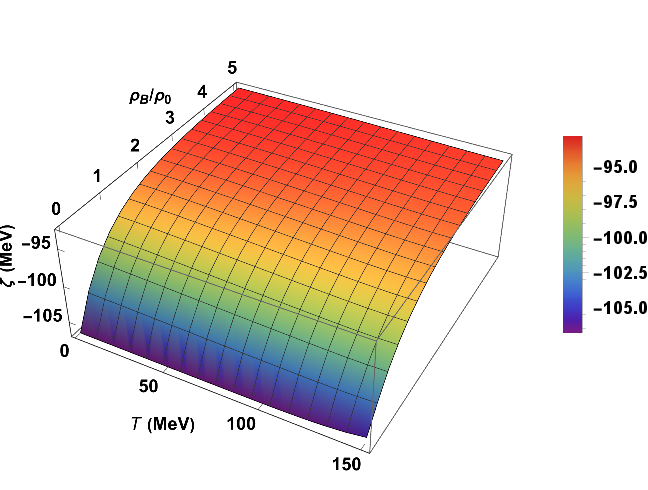}
  \end{subfigure}\hfill % <-- "\hfill"
\begin{subfigure}{.5\linewidth}
  (d)\includegraphics[width=0.9\linewidth]{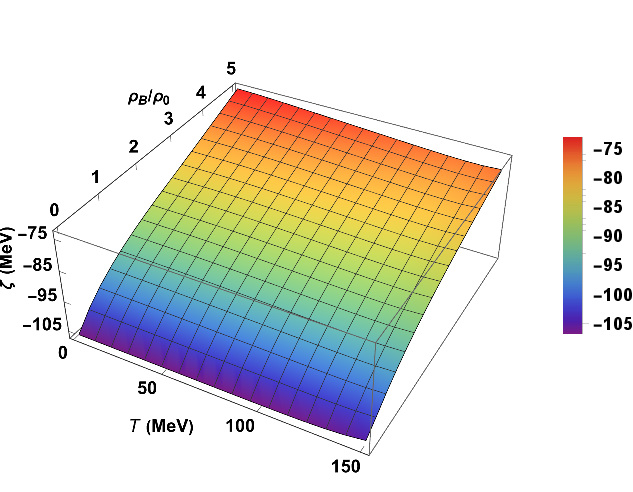}
 \end{subfigure}
 \caption{Same as Fig. \ref {fig3_sigma3D}, for the scalar field $\zeta$.}

\label{fig4_zeta3D}
\end{figure}%%%%%%%
\begin{figure}
    \centering
    \includegraphics[width=0.9\linewidth]{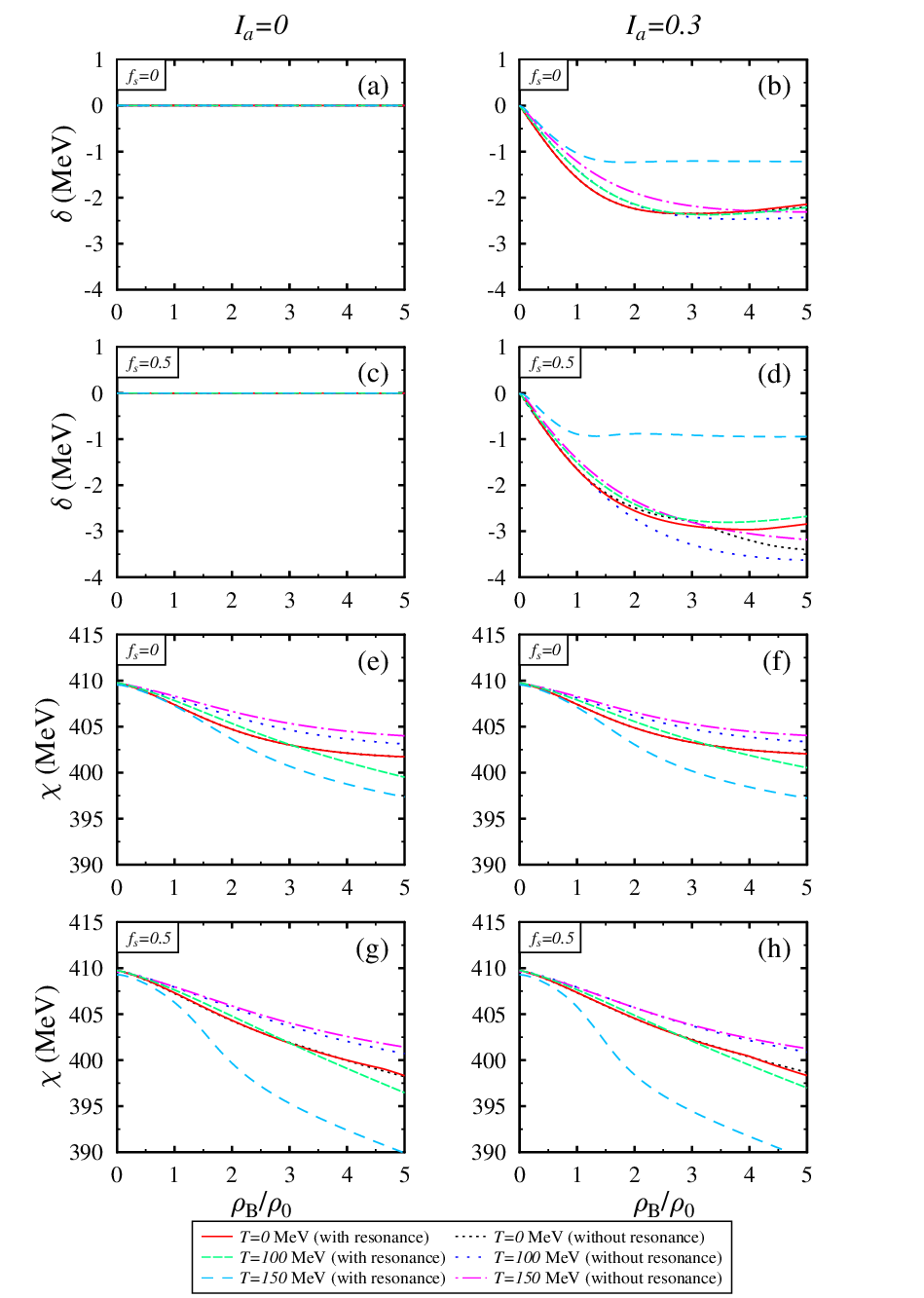}
    \caption{The results of  $\delta$ and $\chi$ fields are plotted as a function of the baryon density ratio $\rho_B/\rho_0$, for isospin symmetric $I_a=0$ and asymmetric $I_a=0.3$ resonance matter. The results are shown for strangeness fractions $f_s = 0$ [in the subplots (a), (b), (e), and (f)] and $f_s = 0.5$ [in subplots (c), (d), (g) and (h)],
 at temperatures, $T = 0, 100$ and $150$ MeV. In each subplot, results are compared with the situation when resonance baryons are not taken into account within the medium.}
\label{fig5_deltachi}
\end{figure}    

\par
From Figs. \ref{fig1_sigmazeta}(a) and (b), we
observe that at temperature $T = 0$ and $f_s =0$,
the values of scalar fields with and without resonance matter ($\Delta$ baryons) appear same for the range of densities shown in these figures.
This is because  at temperature $T = 0$,
$\Delta$ baryons appear at very higher baryon density, whose threshold value depends upon the optical potential $U_\Delta$ considered in the calculations. To understand this, in Fig. \ref{fig2_sigmaUD} we depicted  the scalar field $\sigma$ with respect to the baryonic density for different fixed values of optical potential $U_\Delta$ in the symmetric medium at $T = 0$ and compared the results with case when only nucleons are considered in the calculations. 
%The analysis is conducted at temperature $T=0$ MeV for symmetric matter where $I_a=0$ and $f_s=0$. 
To address the uncertainty in $\Delta$ baryons coupling within dense matter, we utilized the $x_\Delta$ parameter, defined as $x_\Delta = \frac{g_{w\Delta}}{g_{wN}}$, to fix $U_\Delta$. We fitted $x_\Delta$ parameter as 1.0359, 0.9562 ,and 0.8848 corresponding to $U_\Delta=-60$, $-80$, and $-100$ MeV, respectively.
Interestingly, a lower $x_\Delta$ value (0.8848)
% results produced a kink in the
causes difference in value of
 $\sigma$ field in nuclear and $\Delta$ 
 resonance matter at low baryon density ($\rho_B= 4.6\rho_0$), indicating the presence of $\Delta$ baryons beyond that baryon density in the medium.
In the present work, we have not considered the charge neutral matter which is relevant for compact star physics.

\par
\textcolor{blue}{Figs.}  \ref{fig1_sigmazeta} (e) to (h) illustrate how the medium-modified strange scalar-isoscalar field $\zeta$ varies with baryon density ratio $\rho_B/\rho_0$.
% For the given values of $T$, $I_a$, and $f_s$, as baryon density increases, a corresponding decrease in the magnitude of the $\zeta$ field is observed. 
For a fixed value of baryon density $\rho_B$, in the resonance medium, an increase in temperature from $T = 0$ to $150$ MeV causes a decrease in the magnitude of scalar field $\zeta$. This trend is more prominent as baryon density is increased for the fixed values of $T$, $I_a$, and $f_s$. 
It is noticed that the in-medium values of  $\zeta$ field exhibit slightly less drop in their magnitude as one moves from symmetric to asymmetric matter, at fixed strangeness fraction $f_s$. However, increasing the values of $f_s$ at a fixed value of $I_a$, the rate of decrease in the magnitude of $\zeta$ is found to be large. For non-strange medium $f_s=0$, $I_a=0$ (0.3), at  $\rho_B = 4\rho_0$, the magnitude of $\zeta$ field alters by 1$3\% (12.8\%$), $18\% (16.7\%)$, and $21\% (20.7\%)$ and for case of $I_a=0$, $f_s=0$ (0.5), its values modified by $13\% (27.6\%$), $18\% (29\%)$, and $21\% (36.4\%)$ from its vacuum values ($\zeta_0=-106.7$ MeV) at temperatures $T = 0, 100$, and $150$ MeV, respectively. Note that, similar to scalar field $\sigma$, for non-resonance medium, at both $f_s =0$ and $0.5$, the magnitude of scalar field $\zeta$ increases as temperature is changed from $T = 0$ to $150$ MeV.
For more clear visualization of impact of temperature and baryon density of the medium on the scalar fields $\sigma$ and $\zeta$,  $3-D$ plots depicting the variation of scalar fields $\sigma$ and $\zeta$ as a function of $T$ and $\rho_B/\rho_0$ are given in Figs. \ref{fig3_sigma3D} and 
\ref{fig4_zeta3D}.

 \par
  The isospin asymmetry of the medium is considered in the mean-field relativistic models by incorporating the scalar isovector field $\delta$  and is plotted  as a function of the baryon density ratio, $\rho_B/\rho_0$, in subplots $(a)$ to $(d)$ of Fig. \ref{fig5_deltachi}. 
  For the isospin asymmetry $I_a = 0$,
   the value of $\delta$ field  is zero. 
   %This happened due to this field's quark content $\bar {u} u$ -$\bar {d} d$. For $I_a=0$, there is no difference between the scalar density of baryons, hence, no alteration in the magnitude of the $\delta$ field.
   At finite value of isospin asymmetry $I_a$, the magnitude of the $\delta$ field increases as the baryon density
   $\rho_B$ increases from zero to finite value, for certain range.
   For given density of resonance matter,
   the magnitude of $\delta$ field at $T = 150$ MeV is 
   significantly less than the
   values observed at $T = 0$ MeV. 
   This is also consistent with our earlier observations on the behaviour of scalar field at $\eta = 0.3$ and temperature $T = 150$ MeV. As stated earlier, the larger value temperature of the medium  weakens the impact of isospin asymmetry and this effect is more observable in resonance matter.  
   
   % At lower baryonic density, with the rise in temperature, up to T=100 MeV, there is less drop in its value as compared to the higher baryonic density as shown in subplot (b)  of Fig \ref{fig5_deltachi}. The further increase in temperature leads to less drop in the value of the $\delta$ field. These variations are more noticeable when one proceeds from non-strange to strange medium.
\par
In Figs. \ref{fig5_deltachi}(e)-(h), 
the dilaton field $\chi$ is plotted with respect to $\rho_B$ (in units of $\rho_0$). The dilaton field $\chi$ is included in the chiral $SU(3)$ model to simulate the trace anomaly property of QCD.
The in-medium values of scalar dilaton field $\chi$ decrease with increase in the baryon density of the medium. 
For both $f_s = 0$ and $0.5$, the rate of decrease is observed to be more in the
resonance matter. In the isospin symmetric
medium ($I_a = 0$), considering nucleons and $\Delta$ baryons in the medium, at
baryon density $\rho_B = \rho_0 (4\rho_0)$, the values of dilaton field decreases by
$0.58\% (1.86)\%$, $0.47\% (2.1)\%$ and $0.61\% (2.68)\%$ at $T = 0, 100$ and $150$ MeV, respectively, from the vacuum value $\chi_0 = 409.77$ MeV. 
At $f_s = 0.5$, considering other resonance baryons also, the above
values of percentage decrease in the $\chi$ field 
at $\rho_B = \rho_0 (4\rho_0)$, change to $0.59\% (2.38)\%$, $0.52\% (2.6)\%$ and $0.85\% (4.24)\%$ at $T = 0, 100$ and $150$ MeV, respectively. 
Above values of percentage drop can be compared to the situation when resonance baryons are not considered in the calculations. At $f_s = 0$, (having nucleons only), $I_a=0$ and $\rho_B = \rho_0 (4\rho_0)$, values of percentage drop are found to be $0.58\% (1.86)\%$, $0.4\% (1.48)\%$ and $0.35\% (1.28)\%$ $T = 0, 100$ and $150$ MeV, respectively, whereas, in presence of hyperons ($f_s = 0.5$), these
values change to $0.61\% (2.39)\%$, $0.44\% (1.88)\%$ and $0.44\% (1.75)\%$, respectively.
In some past studies of nuclear medium within the chiral SU(3) model, the frozen glueball limit is considered in the calculations due to very small change of
$\chi$ field as a function of baryon density of the medium \cite{Fariborz:2021gtc, Heide:1993yz}.
However, understanding the medium modifications of $\chi$ field has applications in  the study of in-medium properties of charmonium and bottomonium properties through the medium modifications of scalar gluon condensates \cite{Mishra:2014gea, Kumar:2010gb, kumar2011d}. As we know, the non-zero value of trace of energy momentum tensor of QCD is expressed in terms of expectation value of scalar gluon condensates. Within the chiral SU(3) model, the trace of energy momentum tensor is expressed in terms of dilaton field $\chi$ and thus, the in-medium values of scalar gluon condensates can be calculated through the $\chi$ field.
In the QCD sum rules, the properties of charmonium and bottomonium are expressed in terms of scalar gluon condensates and thus,  the medium modifications of these mesons can be studied through the dilaton
field $\chi$.  Since the dilaton field $\chi$ is affected more in the resonance matter as compared to non-resonance case,
 the  medium modifications of the quarkonium properties  in the dense resonance medium may have important consequences and will be explored in future work.

%In this model, the Lagrangian term $\mathcal{L}_{0}$ (Eq.  \ref{eq_L0}) and $\mathcal{L}_{SB}$ (Eq.  \ref{eq_LSB}) , contain the coupling of $\chi$ field with other scalar fields ($\sigma$, $\zeta$, and $\delta$ fields). This coupling lowers the magnitude of the $\chi$ field with isospin asymmetry and strangeness.  

  \subsection{The in-medium masses and optical potentials of $K$ and $\bar K$ mesons}  \label{IIIB}
\begin{figure}
    \centering
    \includegraphics[width=0.9\linewidth, height= 20cm]{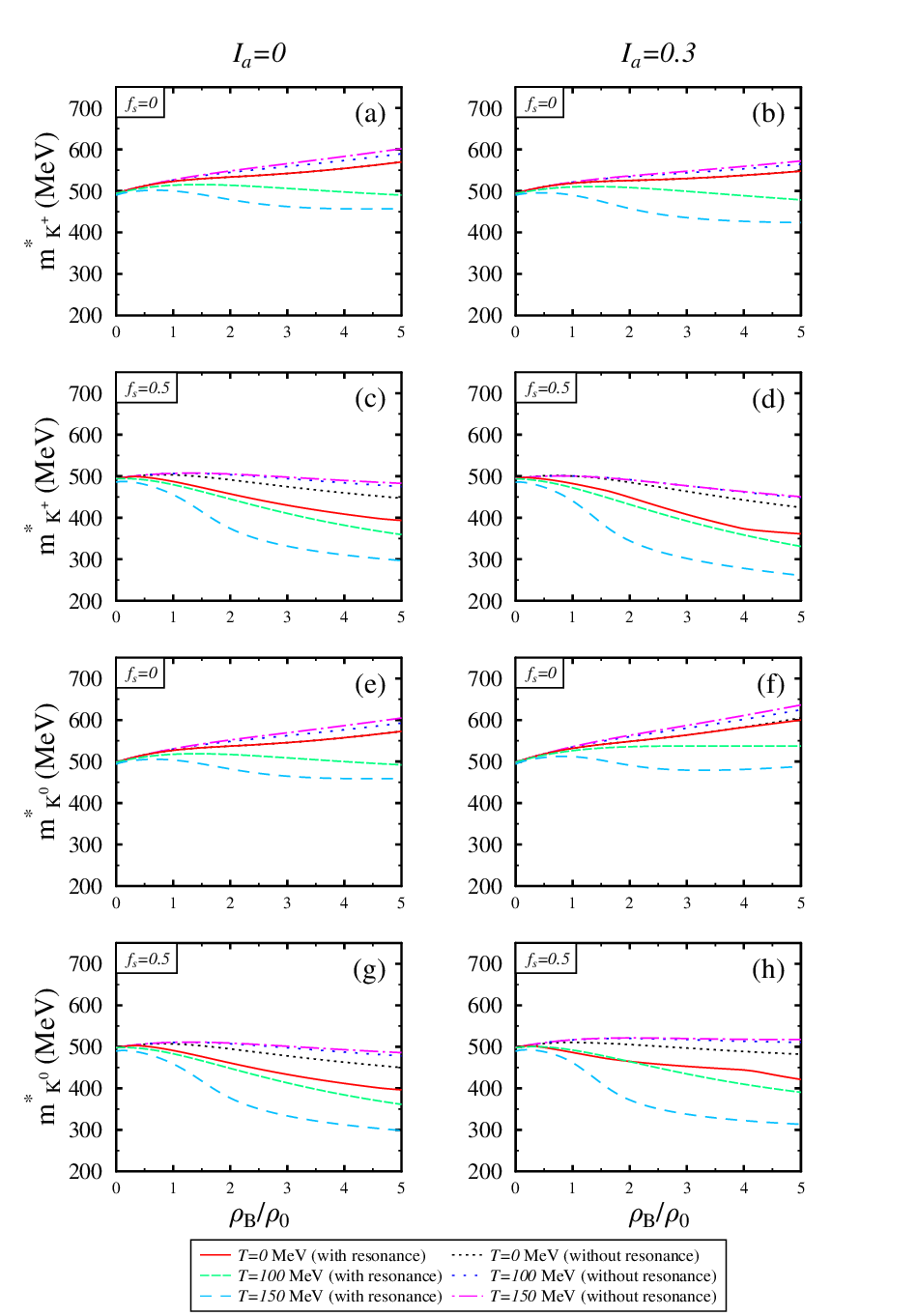}
    \caption{The effective masses for $K^+$ and $K^0$ mesons at zero momentum, plotted as a function of the baryon density ratio $\rho_B/\rho_{0}$ for dense resonance matter. For each value of the isospin asymmetry parameter ($I_a$= 0 and 0.3 ), the results are shown for $f_s = 0$ [in the subplots (a), (b), (e) and (f)] and $f_s = 0.5$ [in subplots (c), (d), (g) and (h)], at temperatures, $T = 0, 100$ and $150$ MeV. These results are further compared in each subplot with the case in which only spin-1/2 baryons are considered. }
    \label{fig6_kpk0}
\end{figure}

\begin{figure} 
    \centering
    \includegraphics[width=0.9\linewidth]{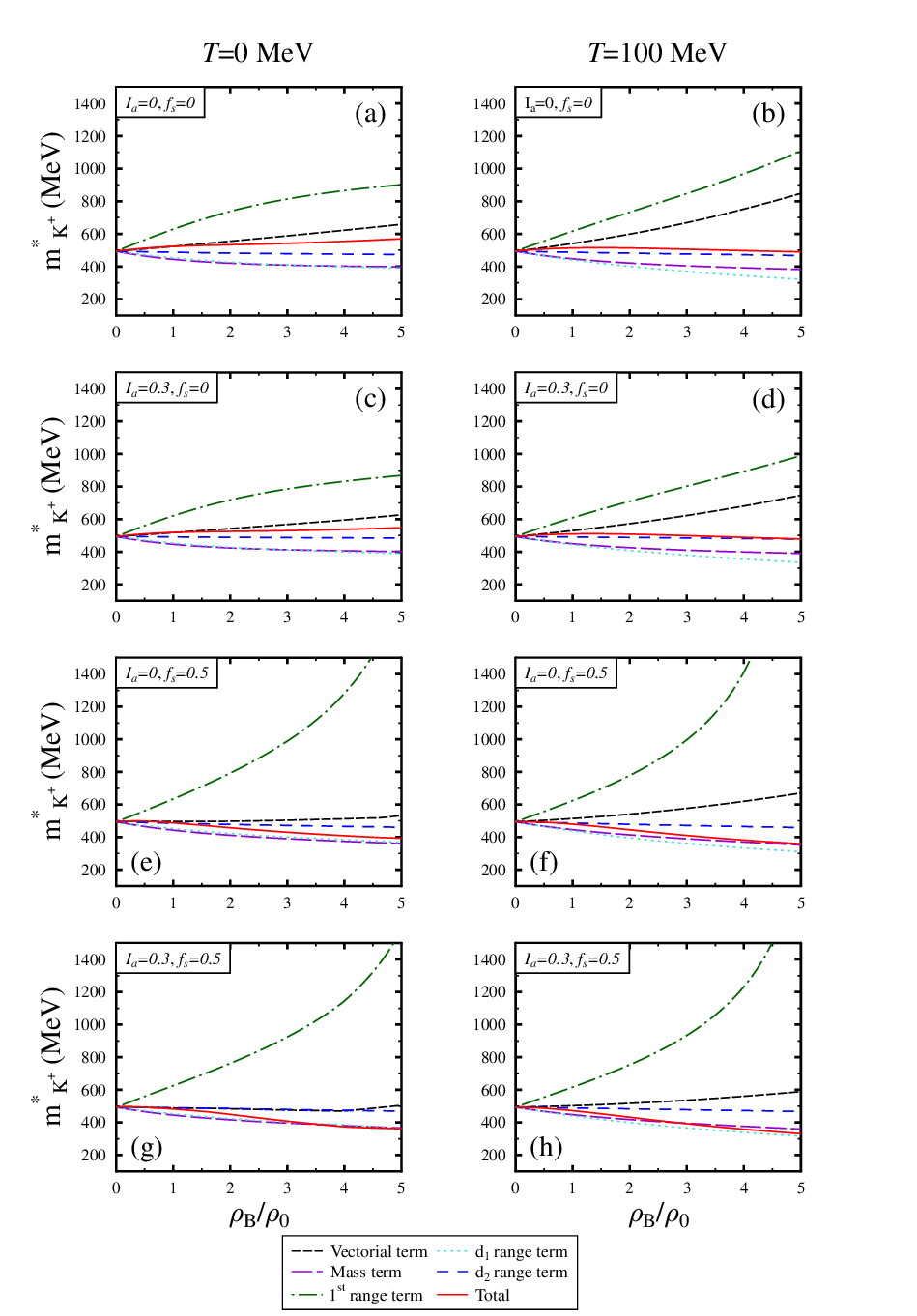}
    \caption{Contributions of individual terms of interaction Lagrangian density to the effective masses of $K^+$ mesons is plotted at temperatures $T=0$ (left panel) and $100$ MeV (right panel), with respect to baryon density ratio $\rho_B/\rho_{0}$, for $I_a = 0,0.3$ and $f_s = 0,0.5$.}
    \label{fig7_kp_terms}
\end{figure}

\begin{figure}
    \centering
    \includegraphics[width=0.9\linewidth]{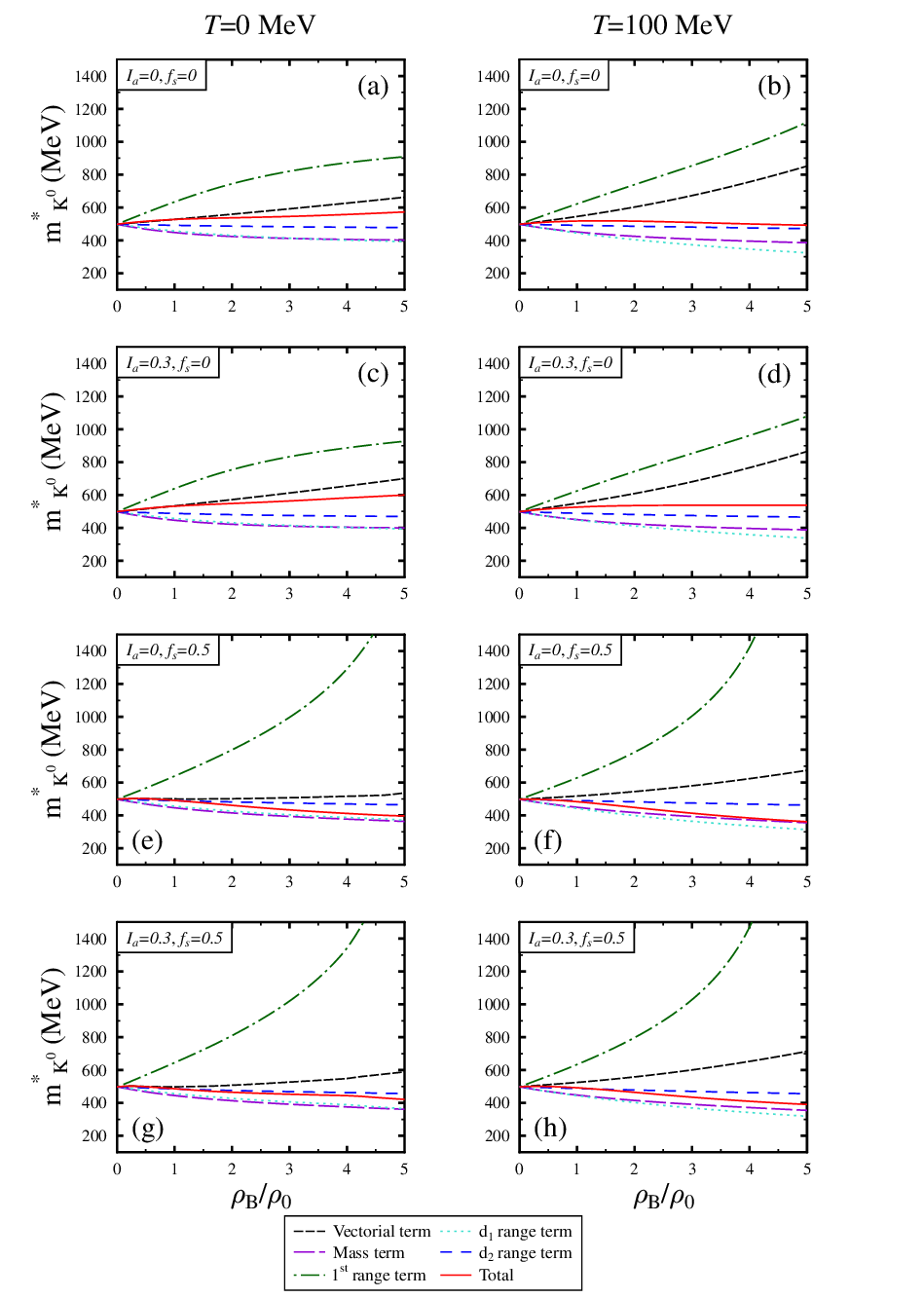}
    \caption{Same as Fig. \ref{fig7_kp_terms}, for $K^0$ meson.
     %mass from various terms of the interaction Lagrangian density plotted as a function of baryon density ratio $\rho_B/\rho_{0}$.
    }
    \label{fig8_k0_terms}
\end{figure}

\begin{figure}[hbt!]
\begin{subfigure}{.5\linewidth}
  (a)\includegraphics[width=0.9\linewidth]{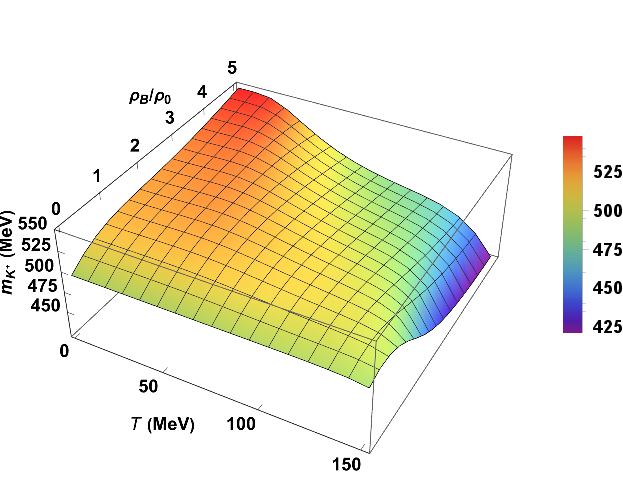}
 
\end{subfigure}\hfill % <-- "\hfill"
\begin{subfigure}{.5\linewidth}
  (b)\includegraphics[width=0.9\linewidth]{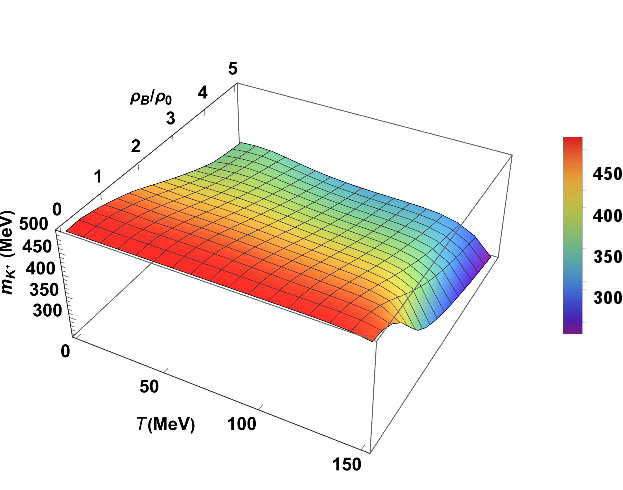}
  \end{subfigure}
\begin{subfigure}{.5\linewidth}
  (c)\includegraphics[width=0.9\linewidth]{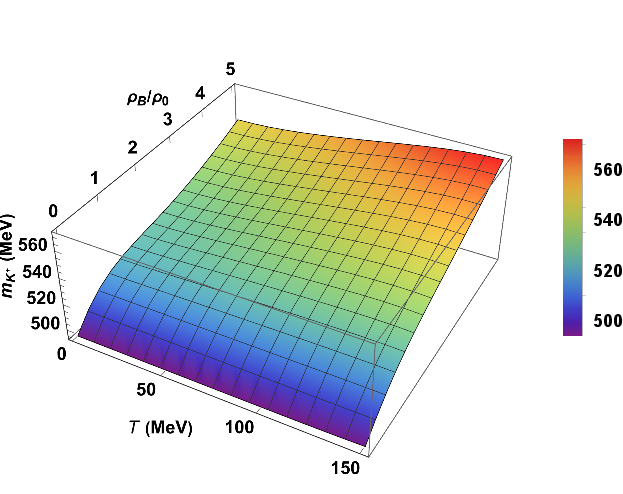}
  \end{subfigure}\hfill % <-- "\hfill"
\begin{subfigure}{.5\linewidth}
  (d)\includegraphics[width=0.9\linewidth]{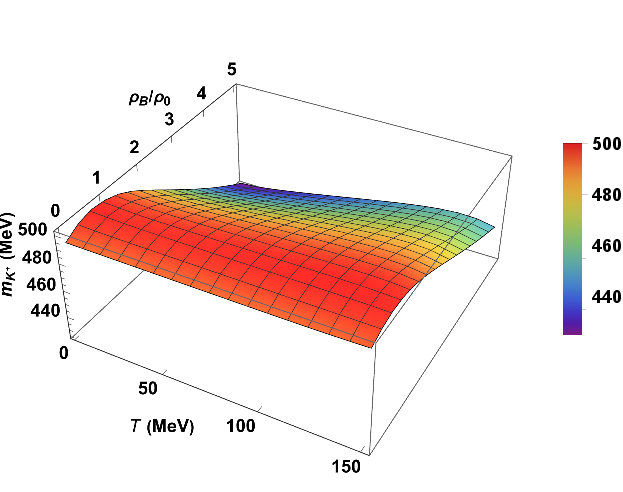}
  \end{subfigure}
\caption{
The 3D representation of effective masses of $K^{+}$ is plotted as a function of baryon density ratio $\rho_B/\rho_0$ and temperature $T$ for isospin asymmetric medium ($I_a=0.3$). The subplots (a) and (c) correspond to $f_s = 0$, with (a) incorporating both nucleons and $\Delta$ baryons while (c) includes only nucleons. The subplots (b) and (d) are for $f_s=0.5$, where (b) consider all octet and decuplet baryons and (d) involves octet baryons only.}
\label{fig9_KP_3D}
\end{figure}

\begin{figure}[hbt!]
\begin{subfigure}{.5\linewidth}
  (a)\includegraphics[width=0.9\linewidth]{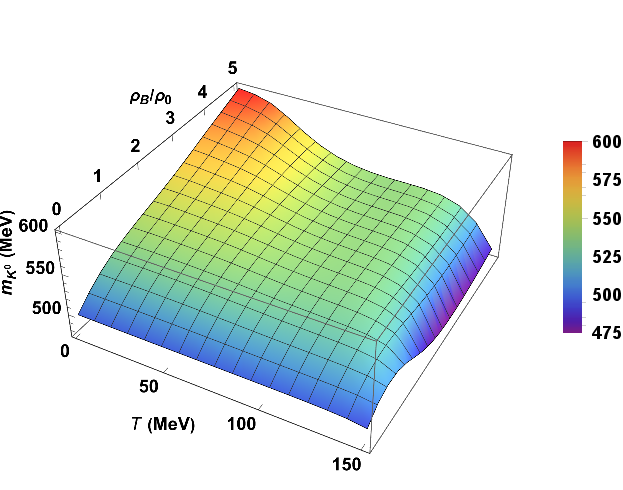}
  % \caption{}
%\label{fig9_KP_3D_a}
\end{subfigure}\hfill % <-- "\hfill"
\begin{subfigure}{.5\linewidth}
 (b) \includegraphics[width=0.9\linewidth]{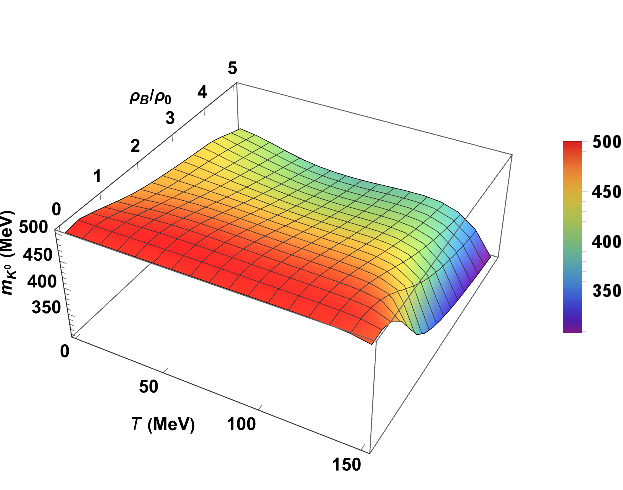}
  % \caption{}
 % \label{fig9_KP_3D_b}
\end{subfigure}
% \medskip % create some *vertical* separation between the graphs
\begin{subfigure}{.5\linewidth}
  (c)\includegraphics[width=0.9\linewidth]{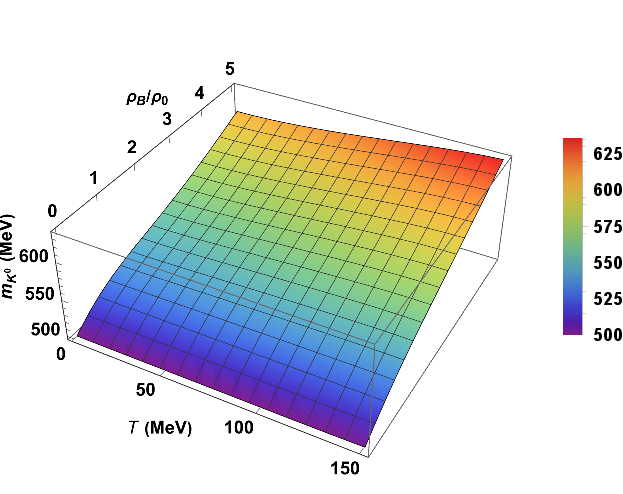}
  % \caption{}
 % \label{fig9_KP_3D_c}
\end{subfigure}\hfill % <-- "\hfill"
\begin{subfigure}{.5\linewidth}
 (d) \includegraphics[width=0.9\linewidth]{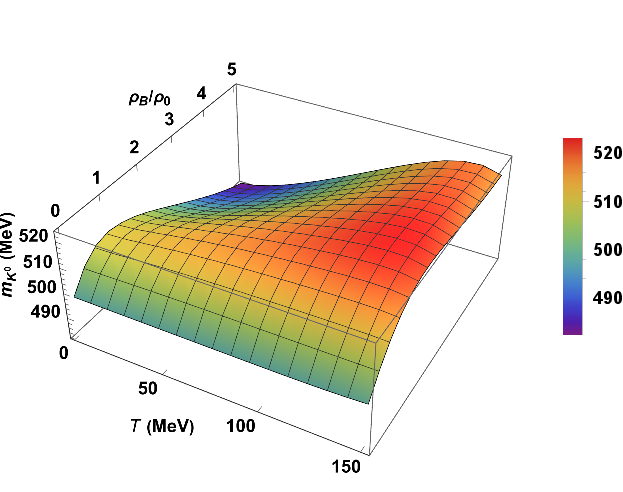}
  % \caption{}
  %\label{fig9_KP_3D}
\end{subfigure}
\caption{Same as Fig. \ref {fig9_KP_3D}, for the effective mass of $K^0$ meson.}
\label{fig10_K0_3D}
\end{figure}

\begin{figure}
    \centering
    \includegraphics[width=0.9\linewidth]{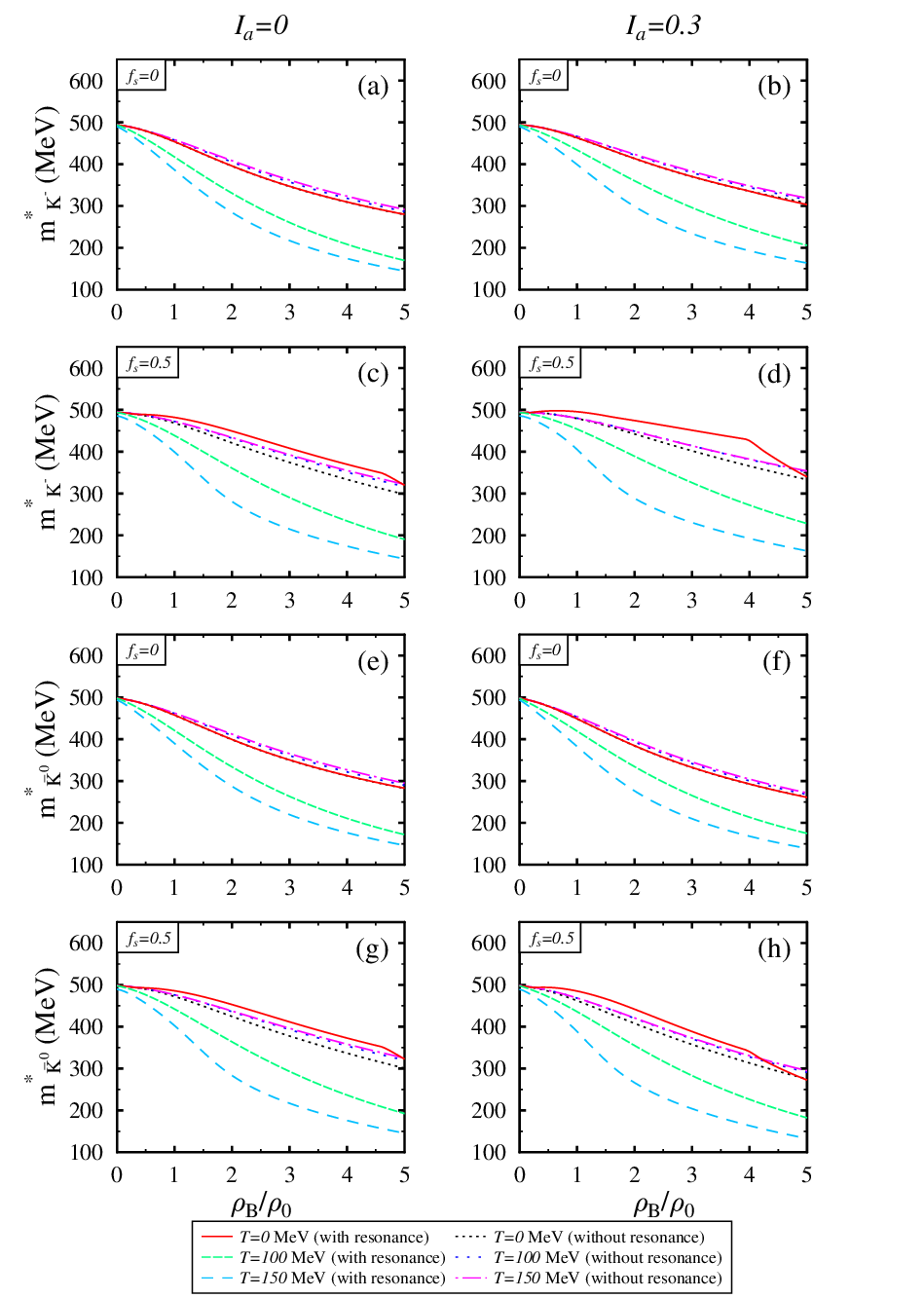}
    \caption{The effective masses for $K^-$ and $\bar K^0$ mesons at zero momentum, plotted as a function of the baryon density ratio $\rho_B/\rho_{0}$  for dense resonance matter. For each value of the isospin asymmetry parameter ($I_a$= 0 and 0.3 ), the results are shown for $f_s = 0$ [in the subplots (a), (b), (e) and (f)] and $f_s = 0.5$ [in subplots (c), (d), (g) and (h)], at temperatures, $T = 0, 100$ and $150$ MeV. These results are further compared in each subplot with the case in which only spin 1/2 baryons are considered.}
    \label{fig11_knk0n}
\end{figure}

\begin{figure}
    \centering
    \includegraphics[width=0.9\linewidth]{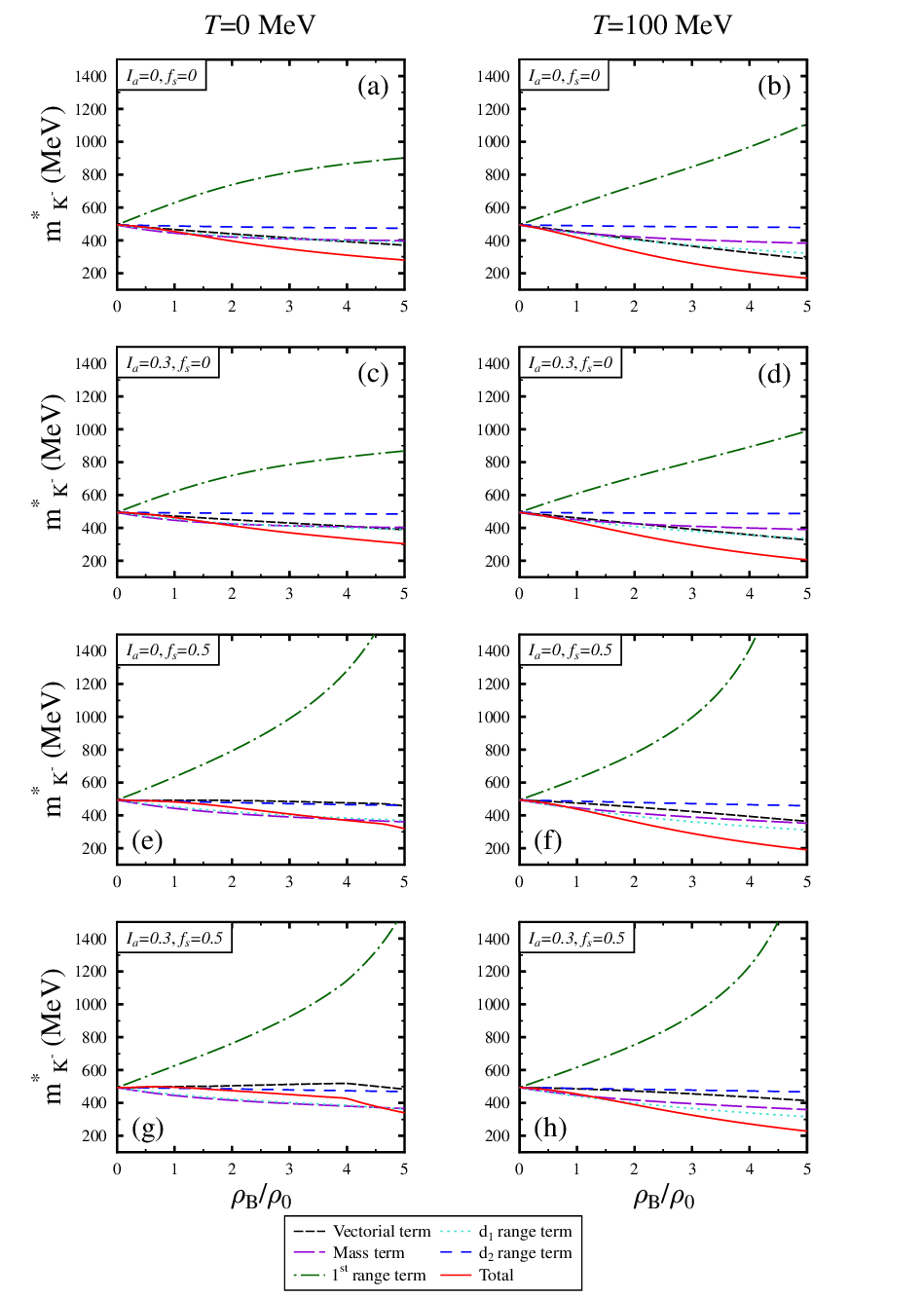}
    \caption{Contributions of individual terms of interaction Lagrangian density to the effective masses of $K^-$ mesons is plotted at temperatures $T=0$ (left panel) and $100$ MeV (right panel), with respect to baryon density ratio $\rho_B/\rho_{0}$, for $I_a = 0,0.3$ and $f_s = 0,0.5$.}
    \label{fig12_kn_terms}
\end{figure}

\begin{figure}
    \centering
    \includegraphics[width=0.9\linewidth]{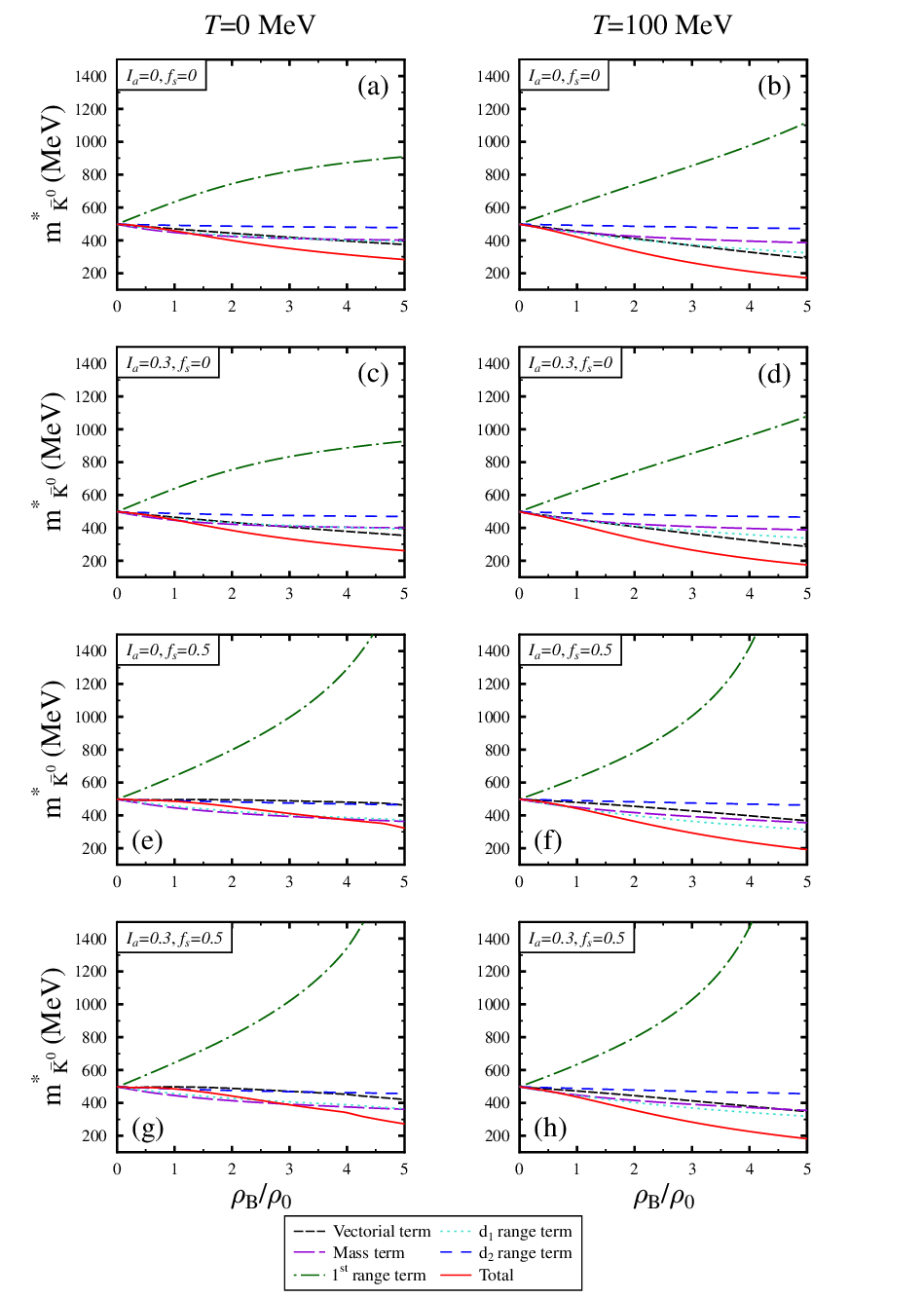}
    \caption{Same as Fig. \ref{fig12_kn_terms}, for $\bar K^0$ mesons.}
    \label{fig13_k0n_terms}
\end{figure}

Now we discuss the in-medium masses and optical potentials of kaons and antikaons in isospin asymmetric dense resonance medium at finite temperature. 
The in-medium properties of $K$  and $\bar K$ mesons were investigated using the chiral SU(3) model in isospin symmetric and asymmetric nuclear and hyperonic matter
in Refs. \cite{Mishra:2004te, Mishra:2006wy, Mishra:2008kg,Mishra:2008dj,kumar2020phi}. This current work considers the impact of dense resonance matter, which comprises nucleons, hyperons, and decuplet baryons, on $K$ and $\bar K$ meson masses.
The effective masses of kaons and antikaons are
calculated by solving the dispersion relation 
given in Eq. (\ref{eq_self_energy}).
%The kaon masses at zero momentum are calculated by solving the Eq.  \ref{eq_self_energy}. 
In Figs. \ref{fig6_kpk0}-\ref{fig13_k0n_terms}, the results of in-medium masses (calculated at zero momentum) of kaons and antikaons are presented against the baryonic density $\rho_B$ for different values of $T$, $I_a$, and $f_s$. The values of effective masses $m^{*}_{K^+}$, $m^{*}_{K^0}$, $m^{*}_{K^-}$, and $m^{*}_{\bar K^0}$ are tabulated in Table \ref{table_mass}. 
In Fig. \ref{fig6_kpk0}
 the variation of in-medium masses of   $K^+$ and $K^0$ mesons is plotted as a function of baryon density ratio $\rho_B/\rho_0$ in dense resonance matter
 for  isospin asymmetry $I_a = 0$ and $0.3$, 
 and strangeness fractions $f_s = 0$ and $0.5$. 
 The results are compared with the situation when
 resonance baryons are not considered in the calculations for temperatures $T = 0, 100$ and $150$ MeV.
 In the nuclear medium, both for zero and finite temperature, the effective masses of 
 $K^{+}$ and $K^{0}$ mesons are observed
 to increase with an increase in the baryon density $\rho_B$, i.e, members of $K$ meson doublet are dominated by repulsive interactions in the nuclear matter. However, when the $\Delta$ baryons are included in the calculations, as can be seen from Figs. 
 \ref{fig6_kpk0} (a), (b), (e) and (f), the rate of increase in the effective mass of $K^{+}$ and $K^{0}$ mesons as a function of baryon density $\rho_B$  of the medium becomes slower, as temperature changes from zero to finite value and for  $T = 150$ MeV, it starts decreasing after certain value of $\rho_B$.
 
 Within the chiral SU(3) model, the effective masses of $K$ and $\bar{K}$ mesons are calculated considering contributions of 
 various interaction terms to the total Lagrangian density as described in Eq. (\ref{Eq_kBD_gen1}). To understand the impact of individual terms to the net effective mass of mesons, in Figs. \ref{fig7_kp_terms} and \ref{fig8_k0_terms} 
 we plotted the contributions of individual terms to effective masses of $K^+$ and $K^{0}$ mesons, respectively, for resonance matter.
In the nuclear medium, the Weinberg Tomozawa term and first range term of the chiral SU(3)
model give repulsive contributions to the effective masses of kaons as a function of density of the medium and dominate over the attractive contributions of scalar terms and $d_1$ and $d_2$ terms  \cite{Mishra:2006wy, Mishra:2008kg, Mishra:2004te}.
However, when the $\Delta$ baryons are included in the calculations, at high temperature, 
contributions due to attractive  terms increases and this causes a slower increase in the mass of kaons as a function of $\rho_B$ and a decrease after certain $\rho_B$ at $T = 150$ MeV.

The finite isospin asymmetry of the medium causes the mass splitting among the members $K^+$ and $K^0$ of $K$ meson isospin doublet. 
For a given density and temperature of the medium, an increase in value of $I_a$ from zero to $0.3$  causes a decrease in the effective mass of $K^{+}$ and an increase in the mass of 
$K^0$ mesons. For example, at $\rho_B = 4\rho_0$, $f_s=0$ and  temperature $T = 0$ MeV, for an increase in $I_a$ from $0$ to $0.3$,
the effective mass of $K^{+}$ ($K^{0}$) decreases (increases) by $16.7 (24)$ MeV, whereas at $T = 100$ MeV these values change to 
$9(37.4)$ MeV. At $T = 100$ MeV, when only nucleons are considered, the above values of mass shift change to $20 (25)$ MeV.
 
%For the non-strange symmetric resonance matter (with nucleons and $\Delta$ baryons), at temperature $T=0$, raising the baryonic density leads to an increase in the masses of $K^+$ and $K^0$ mesons. This is because the vectorial interaction and $1^{st}$ range terms of Eq.  \ref{Eq_kBD_gen1} (previously discussed in Sec. ~\ref{subsec2.2} ) give repulsive contributions to $m^{*}_{K^+}$, $m^{*}_{K^0}$. As can be seen in Fig.  \ref{kp_terms}(a) and \ref{fig8_k0_terms} (a), which shows the contributions to these mesons due to the different terms of interaction Lagrangian, the involvement of these terms suppresses the attractive contribution from the scalar meson exchange, $d_1$, and $d_2$ range terms. 
%\par
%At $T=0$, and $f_s=0$, for the baryonic density $4\rho_0$, the mass of $K^+$ increase by 12\% and 9\% for $I_a=0$ and $0.3$, respectively, with respect to $\rho_B=0$. It indicates that with increasing the value of the isospin asymmetric parameter, there is a slightly less increase in $m^{*}_{K^+}$ as exhibited in the subplots (a) and (c) of Fig.  \ref{kp_terms}. 
\begin{table}
    \centering
    \begin{tabular}{|c|c|c|c|c|c|c|c|c|c|c|c|c|c|}
    \hline
& & \multicolumn{4}{c|}{T=0 MeV}  & \multicolumn{4}{c|}{T=100 MeV}  & \multicolumn{4}{c|}{T=150 MeV} \\
         \cline{3-14}
        &$f_s$ & \multicolumn{2}{c|}{$I_a=0$} & \multicolumn{2}{c|}{$I_a=0.3$} & \multicolumn{2}{c|}{$I_a=0$} & \multicolumn{2}{c|}{$I_a=0.3$} & \multicolumn{2}{c|}{$I_a=0$} & \multicolumn{2}{c|}{$I_a=0.3$} \\
        \cline{3-14}
         & &$\rho_0$&$4\rho_0$ &$\rho_0$&$4\rho_0$ &$\rho_0$&$4\rho_0$ &$\rho_0$&$4\rho_0$ &$\rho_0$&$4\rho_0$ &$\rho_0$&$4\rho_0$ \\ \hline
	$m^{*}_{K^+}$&0&523&554.3&518.6&537.6&514.6&497.4&510&488.5&500&457&489.6&426.8\\
		\cline{2-14}
		&0.5&487.2&408.6&482&373.4&480&381.8&472&358.4&455&310.3&440.4&278\\
		\cline{1-14}
		
		$m^{*}_{K^0}$&
		0&527&557.5&531.8&582.3&517.4&500&526.2&537.4&503&459&511.8&481\\
		\cline{2-14}
		&0.5&491&411.8&486.4&444.2&483&384&499.2&410&458.8&312&461.8&322.3\\
		\cline{1-14}
		$m^{*}_{K^-}$&
		0&453.7&309.6&463&335.4&417&208&434&245.4&386.4&174&399.2&193\\
		\cline{2-14}
		&0.5&482&370&495.6&426.5&438.6&234&454&272&399.2&174.1&405.4&192.4\\
		\cline{1-14}
		
		$m^{*}_{\bar K^0}$&
		0&457.7&313&449&293&421&210&419&213&390&176&382.7&168\\
		\cline{2-14}
		&0.5&486&373&485.2&340.4&442.3&236&436&226.2&402.6&175.9&388.7&163.4\\
		\cline{1-14}

    \end{tabular}
    
    \caption{The in-medium values (in the units of MeV) of $m^{*}_{K}$  and $m^{*}_{\bar K}$ at different fixed values of $\rho_B$, $I_a$, $f_s$, and $T$  }
    \label{table_mass}
\end{table}


%%%%%%%%%%%%%%%KP
%%%%%%%%%%%%%%%%%%%%%%%%%%%%

% For non-zero values of $f_s$, masses of $K^+$ meson decrease with increase in the values of $\rho_B$ and $T$ at a fixed value of $I_a$.

In the subplots $(c), (d), (g)$ and $(h)$ of Fig. \ref{fig6_kpk0},
the effective masses of $K^{+}$ and $K^{0}$ mesons are plotted for strangeness fraction $f_s = 0.5$
in the resonance medium, considering the presence of all decuplet baryons  along with nucleon and hyperons.
In Ref. \cite{kumar2020phi, Mishra:2008dj}, the impact of hyperons 
along with nucleons was studied on the effective masses of kaons and antikaons.
The presence of hyperons in the medium is observed to cause 
a decrease in the effective mass of kaons due to dominant attractive interactions (see lines without resonances in the subplot for $f_s = 0.5$ of Fig. \ref{fig6_kpk0}).
As can be seen from the results, when the decuplet baryons having strange quark content are included in the calculations along with nucleons and hyperons, the effective masses of $K^{+}$ and $K^{0}$ mesons decreases more rapidly as a function of $\rho_B$ and rate of 
decrease becomes more at high value of $T$.
For example, in isospin asymmetric medium with $I_a = 0.3$,  at baryon density $\rho_B = \rho_0 (4\rho_0)$, the mass shift for $K^{+}$
mesons is observed to be $-53.6 (-216)$ MeV in the presence of decuplet baryons, whereas with nucleons and hyperons only, these values of mass shift change to $-6.5 (-31)$ MeV, at $T = 150$ MeV.
For $K^{0}$ mesons, the above values of mass shift are found to be 
$-36.2 (-175.7)$ MeV and $-19 (-20)$ MeV, with and without decuplet baryons, respectively.
The drop in the masses of $K$ mesons as a function of baryon density of the strange medium is due to the decrease in the repulsive contributions from the vectorial Weinberg Tomozawa term and increase in the attractive contributions from scalar meson exchange (mass term) and $d_1$ and $d_2$ terms.
In Figs.~\ref{fig9_KP_3D} and \ref{fig10_K0_3D}, we have also shown the  $3D$ plot of  the effective masses of kaons, $K^+$ and $K^0$, respectively, as a function of baryon density ratio $\rho_B/\rho_0$ and temperature $T$, for $f_s = 0$ and $0.5$, respectively, keeping isospin asymmetry $I_a$ fixed at $0.3$.

%With the further rise in strangeness fraction, the value of $m^{*}_{K^+}$  significantly drops in the resonance medium for the higher density relative to the lower one, as shown in Fig.  \ref{fig6_kpk0} (c) and (d). For example, at $f_s=0.5$ and $\rho_B$= $4\rho$, the masses of $K^+$ are 408.6 (373.4), 396 (370), and 324 (288.7) MeV at temperatures $T=0$, $100$, and $150$ MeV respectively for $I_a=0$ (0.3) as exhibits in Table \ref{table_mass}.

\par
%The in-medium mass of $K^0$ mesons at zero momentum shows a comparable pattern to that of $K^+$ mesons (which is expected) as both mesons are members of the same isospin doublet. At $f_s=0$, $\rho_B= \rho_0$ and temperature $T=100$ MeV, the drop in  $m^{*}_{K^+}$  is 5 Mev and an increase in  $m^{*}_{K^0}$ is 8.5 MeV as we shift from $I_a=0$ to $0.3$. The combined contribution of vectorial interaction,  $\delta$ meson, and isospin-dependent $d_2$ range terms lead to their opposite response on isospin asymmetric parameter. The transition from $f_s=0$ to 0.5 gives the decrement in the values of $m^{*}_{K^+}$ and $m^{*}_{K^0}$, which is 34 Mev and 36 MeV respectively at $I_a=0$, $\rho_B= \rho_0$ and temperature $T=100$ MeV. %This drop in their values becomes more notable at higher baryonic densities. 
\begin{figure}[hbt!]
\begin{subfigure}{.5\linewidth}
 (a)\includegraphics[width=0.9\linewidth]{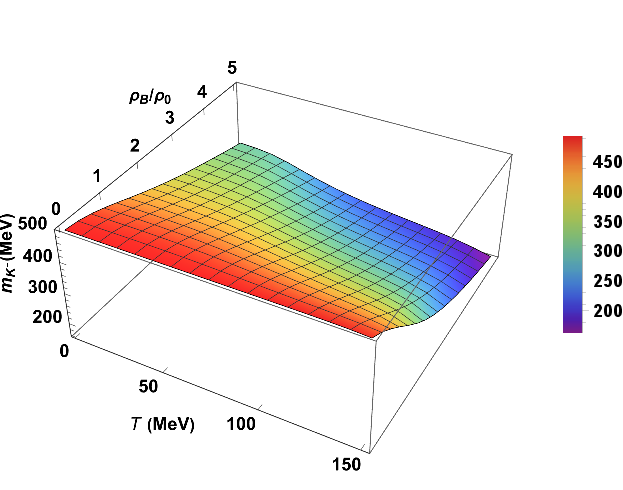}
  % \caption{}
%\label{fig9_KP_3D_a}
\end{subfigure}\hfill % <-- "\hfill"
\begin{subfigure}{.5\linewidth}
  (b)\includegraphics[width=0.9\linewidth]{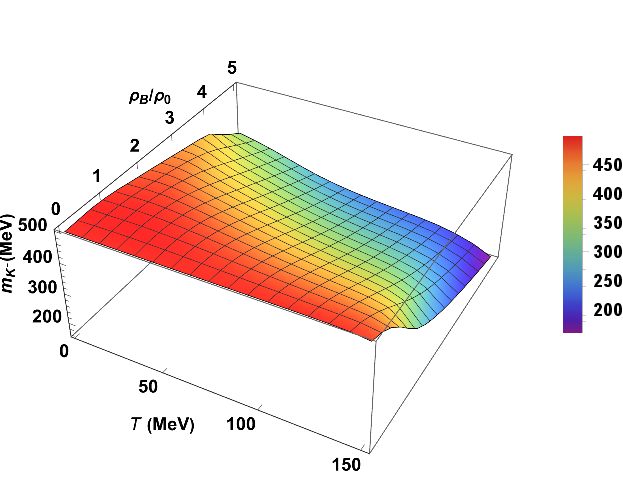}
  % \caption{}
 % \label{fig9_KP_3D_b}
\end{subfigure}
% \medskip % create some *vertical* separation between the graphs
\begin{subfigure}{.5\linewidth}
  (c)\includegraphics[width=0.9\linewidth]{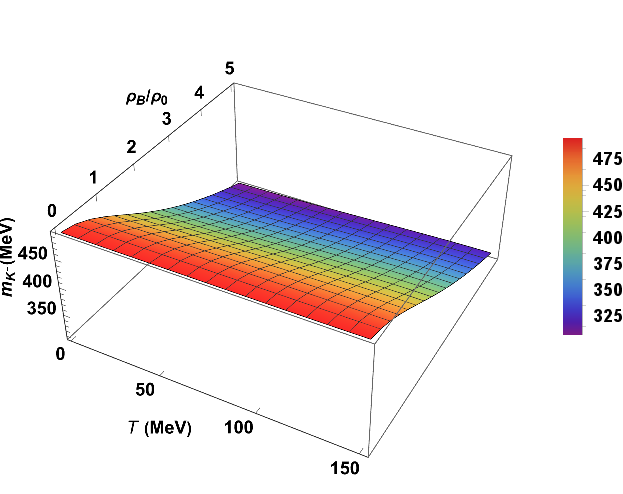}
  % \caption{}
 % \label{fig9_KP_3D_c}
\end{subfigure}\hfill % <-- "\hfill"
\begin{subfigure}{.5\linewidth}
  (d)\includegraphics[width=0.9\linewidth]{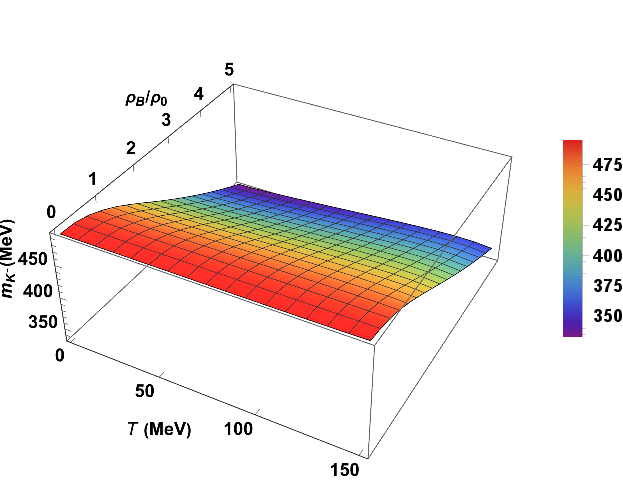}
  % \caption{}
  %\label{fig9_KP_3D}
\end{subfigure}
\caption{Same as Fig. \ref {fig9_KP_3D}, for the effective mass of $K^-$ meson.}
\label{fig14_KN_3D}
\end{figure}

\begin{figure}[hbt!]
\begin{subfigure}{.5\linewidth}
  (a)\includegraphics[width=0.9\linewidth]{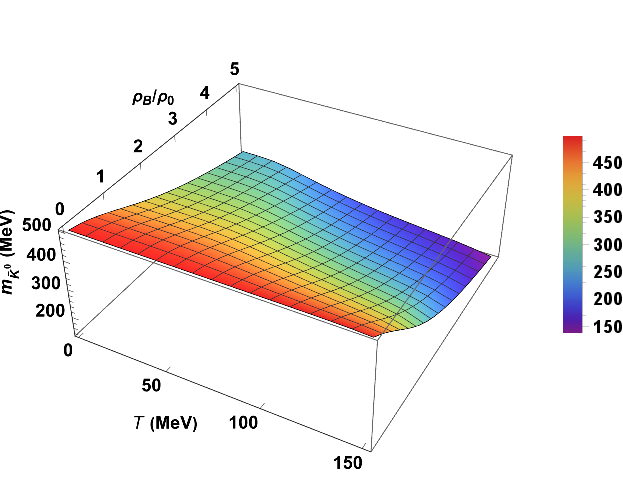}
  % \caption{}
%\label{fig9_KP_3D_a}
\end{subfigure}\hfill % <-- "\hfill"
\begin{subfigure}{.5\linewidth}
  (b)\includegraphics[width=0.9\linewidth]{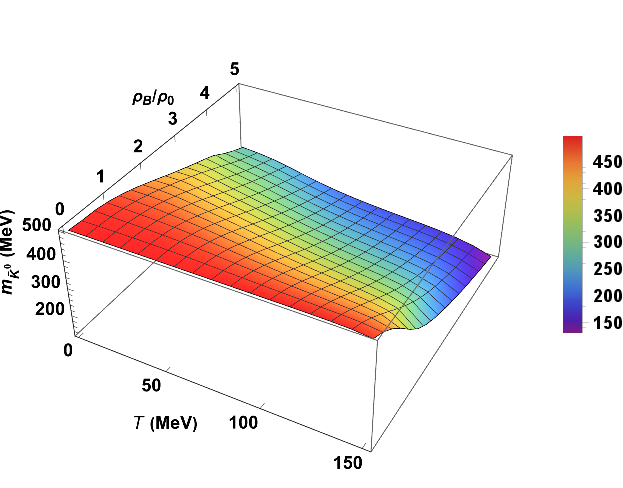}
  % \caption{}
 % \label{fig9_KP_3D_b}
\end{subfigure}
% \medskip % create some *vertical* separation between the graphs
\begin{subfigure}{.5\linewidth}
  (c)\includegraphics[width=0.9\linewidth]{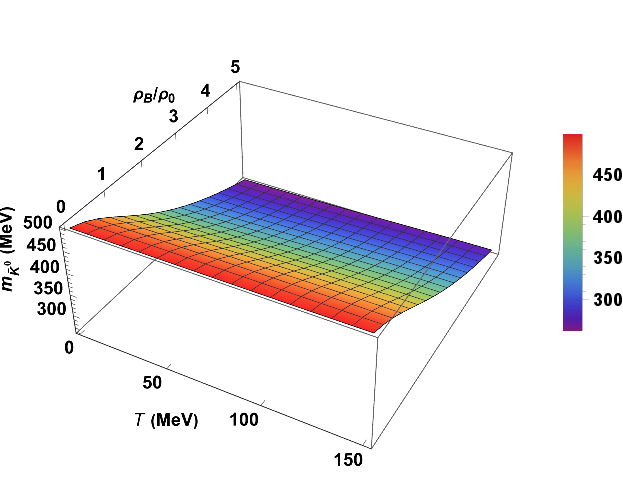}
  % \caption{}
 % \label{fig9_KP_3D_c}
\end{subfigure}\hfill % <-- "\hfill"
\begin{subfigure}{.5\linewidth}
  (d)\includegraphics[width=0.9\linewidth]{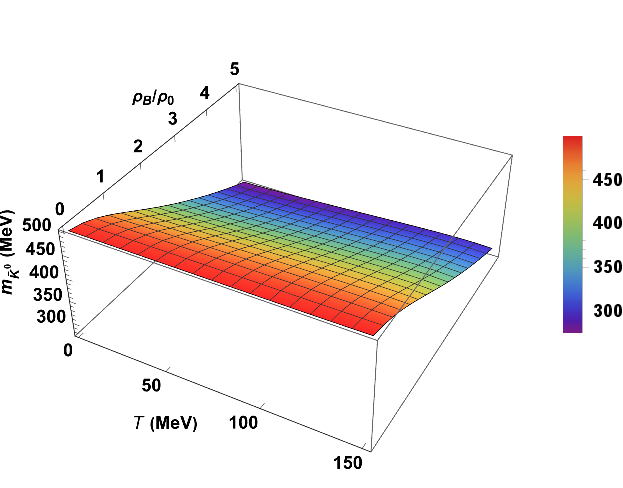}
  % \caption{}
  %\label{fig9_KP_3D}
\end{subfigure}
\caption{Same as Fig. \ref {fig9_KP_3D}, for the effective mass of $\bar K^0$.}
\label{fig15_K0N_3D}
\end{figure}
\par
Now we examine  the effective masses of antikaons $K^-$ and $\bar K^0$ belonging to $\bar{K}$ isospin doublet  in the resonance matter. In Figs. \ref{fig11_knk0n} we plotted the effective masses of $K^-$ and $\bar K^0$ mesons
as a function of baryon density $\rho_B$ (in units of $\rho_0$) in the dense resonance matter
and compared the results with non-resonance case.
 Figures \ref{fig12_kn_terms} and \ref{fig13_k0n_terms}
 illustrate the contributions of individual terms of the Lagrangian density
 to the effective masses of $K^-$ and $\bar{K^0}$ mesons, respectively.
 
 %%check
 For a given temperature, isospin asymmetry and strangeness fraction of the medium, the in-medium masses of $K^-$  and $\bar K^0$ mesons are observed to decrease with an increase in the baryon density $\rho_B$. 
 Recall that in the  non strange resonance medium ($f_s = 0$), at temperature $T=0$ MeV, the effective masses of kaons, $m^{*}_{K^+}$ and $m^{*}_{K^0}$, were increasing with the baryon density of the medium.
% as depicted in Fig.  \ref{fig11_knk0n}. 
 This  difference in the in-medium modification of kaons and antikaons is due to the vectorial interaction term which offers an attractive contribution to antikaons but is repulsive for kaons. 
 Finite isospin asymmetry in the medium also causes the splitting in the effective masses of $K^-$ and $\bar K^0$
 mesons. For instance, at nuclear saturation nuclear density $\rho_B = \rho_0$, temperature  $T=0$ MeV and strangeness fraction $f_s=0$, the effective mass $m^{*}_{K^-}$ increases by 9.3 MeV and  $m^{*}_{\bar K^0}$  decreases by  8.7 MeV, as isospin asymmetry parameter $I_a$ is changed from $0$ to $0.3$.
 As can be seen from subplots (c), (d), (g) and (h) of Fig. \ref{fig11_knk0n}, for zero temperature and at strangeness fraction $f_s = 0.5$, considering all decuplet baryons, the effective mass of
 antikaons is observed to be more 
 as compared case when these decuplet baryons are not taken into account.
 For a given value of $\rho_B$, $I_a$ and $f_s$, in the resonance medium, the masses of antikaons are observed to be smaller at $T = 100$ and $150$ MeV than $T =0$ MeV.
  For example, in isospin asymmetric medium and $f_s = 0$, at baryon density $\rho_B = \rho_0 (4\rho_0)$, the effective masses $m_{K^{-}}^{*}$ and $m_{\bar{K^{0}}}^{*}$ decrease
 by 63.8 (142.4) and 66.3 (125)  MeV, respectively, as temperature $T$ is changed from $T = 0$ to 150 MeV.
At $f_s = 0.5$, these values change to
90.2 (234)  and 96.5 (177)  MeV, respectively.
 However, as can be seen from Fig. \ref{fig11_knk0n}, without resonances,
 the masses of antikaons were found to be
 more at  $T = 100$ and $150$ MeV as compared to $T =0$ situation.
 For example, for $f_s = 0.5$ and $I_a = 0.3$, the effective masses  $m_{K^{-}}^{*}$ and $m_{\bar{K^{0}}}^{*}$ increase by 0.5 (16.62) and 6.32 (17.64)  MeV at $\rho_0 (4\rho_0)$, as $T$ is changed from zero to $150$ MeV.
 In Figs.~\ref{fig14_KN_3D} and \ref{fig15_K0N_3D}, we have shown $3D$ plot of  the effective masses of antikaons $K^-$ and $\bar{K^0}$ mesons, respectively, as a function of baryon density and temperature, for $f_s = 0$ and $0.5$, and isospin asymmetry $I_a = 0.3$.
 % Within a strange resonance medium ($f_s=0.5$), the rate of drop of antikaon masses increases compared to a non-strange medium.
\begin{figure}
    \centering
    \includegraphics[width=0.9\linewidth]{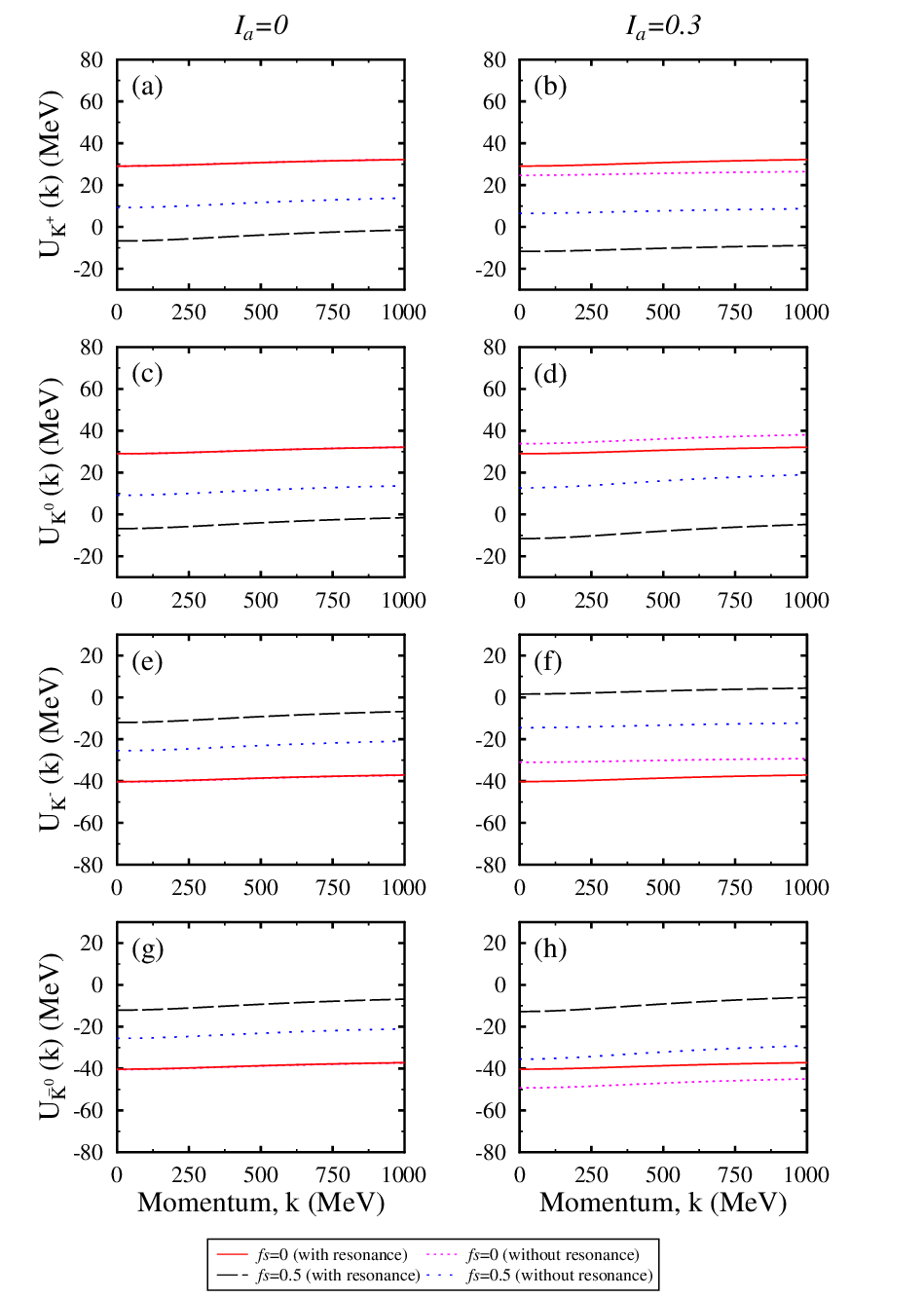}
    \caption{The optical potential for kaons and antikaons [ for $ K^{+}$ in (a) and (b), $ K^{0}$ in (c) and (d), $ K^{-}$ in (e) and (f), and $\bar K^{0}$ in (g) and (f) ] in MeV, plotted as a function of momentum at $\rho_B$=$\rho_{0}$ and temperature $T=0$ MeV, for  $I_a$ = 0 and 0.3. For each value of $I_a$, the optical potentials are shown for $f_s$=0 (solid lines, representing nucleons and $\Delta$ baryons in the medium) and $f_s$=0.5 (dashed lines, representing all resonance baryons). In each subplot, dotted lines correspond to the medium without considering resonance baryons.} 
    \label{fig16_op_rho0}
\end{figure}
\begin{figure}
    \centering
    \includegraphics[width=0.9\linewidth]{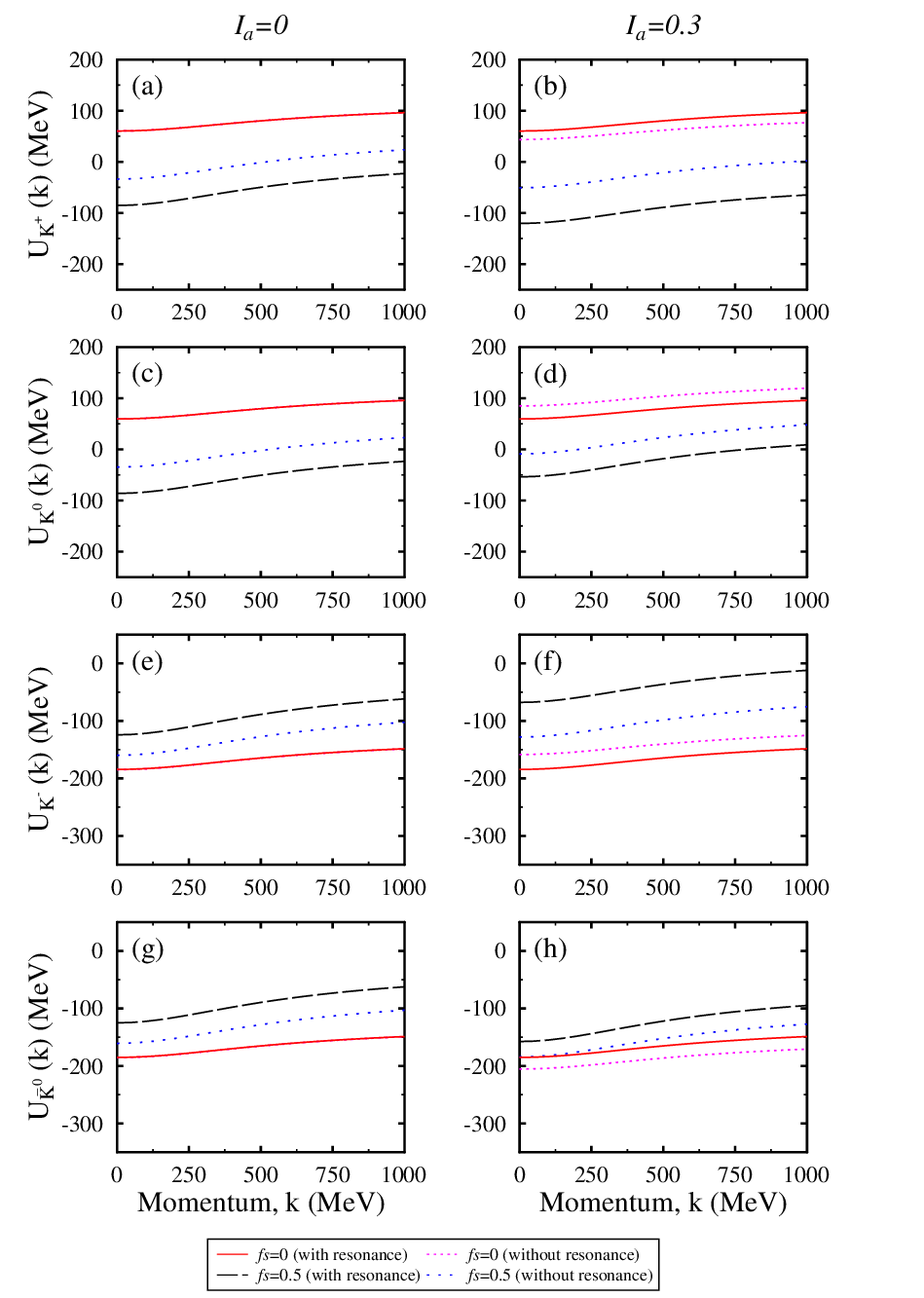}
    \caption{Same as Fig. \ref {fig16_op_rho0}, for $\rho_B$=$4\rho_{0}$ and $T=0$ MeV}

    \label{fig17_op_4rho0}
\end{figure}
%\par 
%At $f_s=0.5$ and $T=0$ MeV, a steep decline appears in the $m^{*}_{K^-}$ ($m^{*}_{\bar K^0}$) when baryonic density exceeds the $4\rho_0$, as shown in Fig.  \ref{knk0n} (c), (d), (g) and (h). It can be attributed to (i) $\Delta$ baryons start populated in the medium, (ii) the isospin symmetric $d_1$ range term becomes more negative, and (iii) similar kink observed in vectorial interaction term at that particular density, as depicted on \textcolor{blue}{Figs.}  \ref{kn_terms} and \ref{k0n_terms}. As temperature rises, the values of effective masses of antikaons are reduced as the function of $\rho_B$, and this reduction becomes even more remarkable for strange medium. For example, at $T=150$ MeV, $f_s=0.5$, when $\rho_B$ changes from $\rho_0$ to $\rho_B$, $m^{*}_{K^-}$ ($m^{*}_{\bar K^0}$) decreased by 226 (228) MeV. From these results, it can be inferred that antikaons are more sensitive to temperature changes compared to variations in strangeness fraction and isospin asymmetry. In each subplot of Fig. \ref{knk0n}, the masses of $K^-$ ($\bar K^0$) for the non-resonance medium are also plotted. In this medium, $m^{*}_{\bar K^0}$ and $m^{*}_{K^-}$ exhibits more dependency on $f_s$ as compared to $I_a$ and $T$.
  
\begin{figure}
    \centering
    \includegraphics[width=0.9\linewidth]{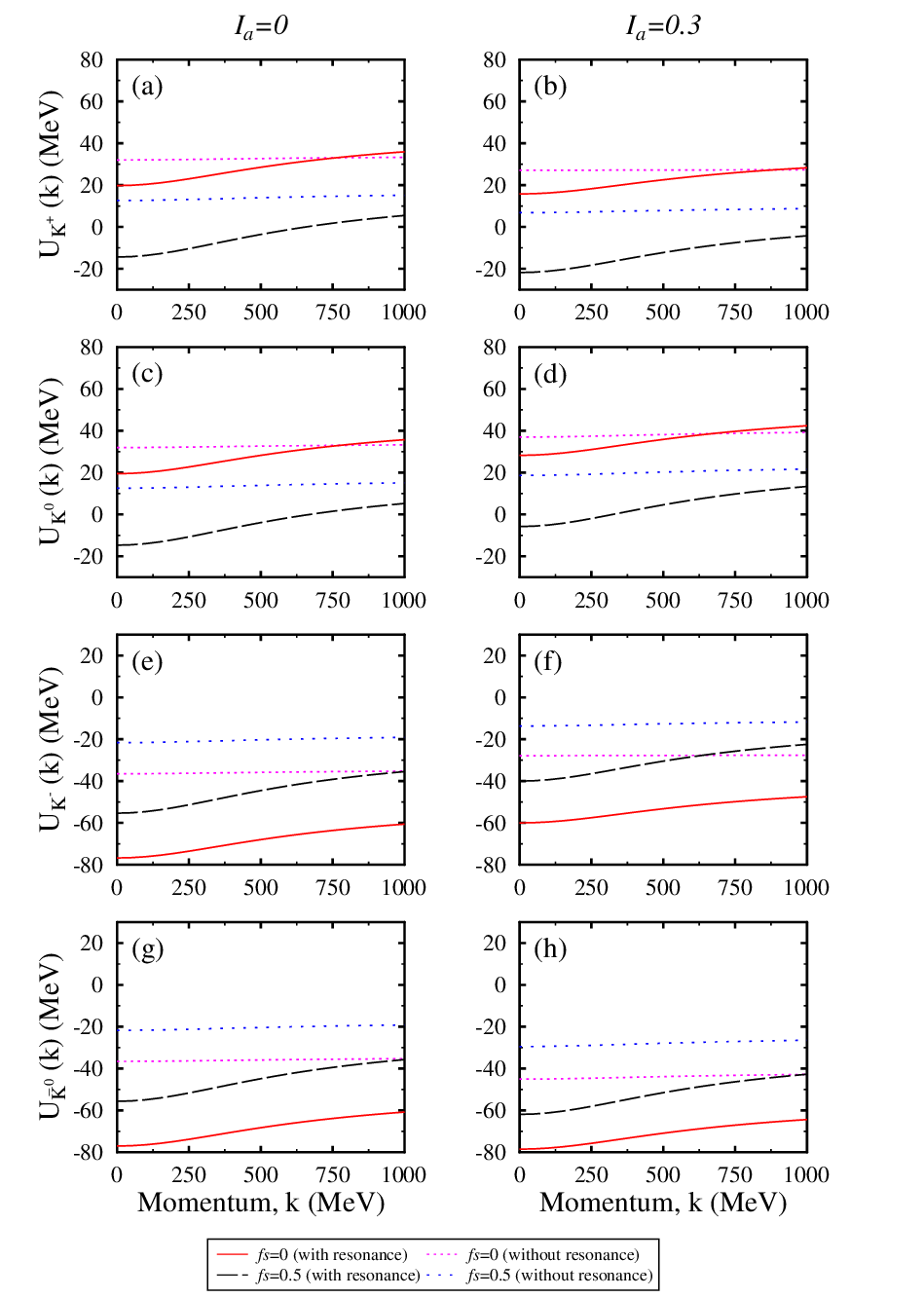}
    \caption{The optical potential for kaons and antikaons depicted as a function of the momentum (in MeV) at $\rho_B$=$\rho_{0}$ and $T=100$ MeV, for $I_a$ = 0 and 0.3. This figure shows optical potential of $ K^{+}$ in (a) and (b), $ K^{0}$ in (c) and (d), $ K^{-}$ in (e) and (f), and $\bar K^{0}$ in (g) and (f) ]. For each $I_a$ value, optical potential are shown for $f_s$=0 (solid lines, representing nucleons and $\Delta$ baryons in the medium) and $f_s$=0.5 (dashed lines, indicating all resonance baryons). In each subplot, dotted lines correspond to the medium without considering resonance baryons.} 
    \label{fig18_opT_rho0}
\end{figure}
\begin{figure}
    \centering
    \includegraphics[width=0.9\linewidth]{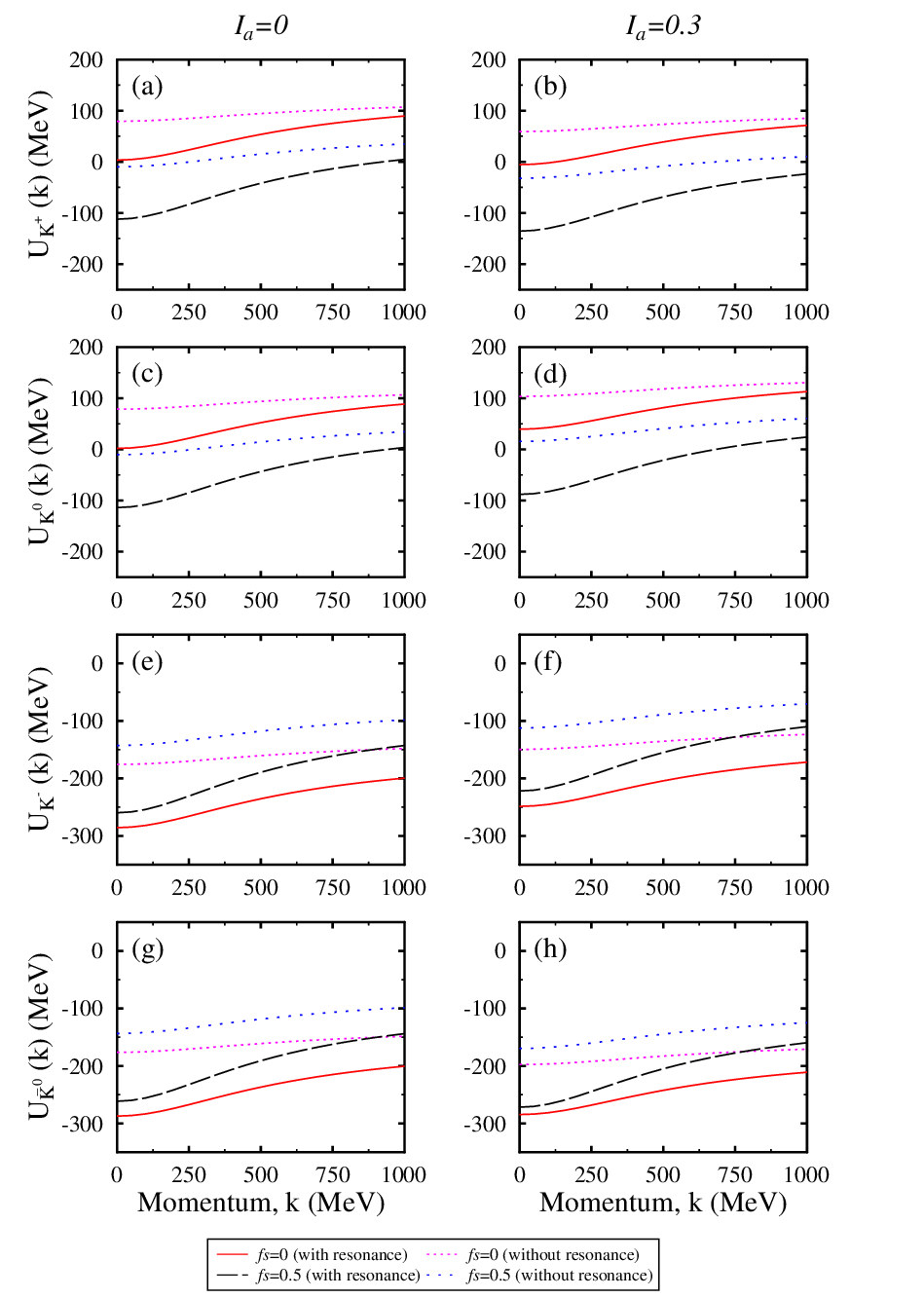}
    \caption{Same as Fig. \ref {fig18_opT_rho0}, for $\rho_B$=$4\rho_{0}$ and $T=100$ MeV}
    \label{fig19_opT_4rho0}
\end{figure}

\par 
The optical potentials of kaons and antikaons at finite momentum can be calculated using Eq.~\ref{Eq_optK}.
    In Figs. \ref{fig16_op_rho0} and \ref{fig17_op_4rho0} the optical potentials of kaons $(K^{+}, K^{0})$ and antikaons $(K^{-}, \bar{K^{0}})$ are plotted as a function of momentum $k$ for baryon density $\rho_0$ and $4\rho_0$, respectively and keeping temperature $T = 0$ MeV, whereas the Figs.  \ref{fig18_opT_rho0} and  \ref{fig19_opT_4rho0}
    show the results at temperature $T = 100$ MeV. For a given baryon density of the medium, the depth of optical potentials of kaons and antikaons decreases as the value of momentum $k$ is increased. The impact of momentum is more visible at high baryon densities.
  For example, considering all decuplet baryons along with nucleons and hyperons in the medium, at $f_s = 0.5$, $\rho_B =\rho_0$, $I_a = 0.3$ and $T = 100$ MeV, as the value of momentum $k$ is increased from zero to 
  $500$ MeV, the values of optical potentials for $K^{+}, K^{0}, K^{-}$ and $\bar{K^0}$ mesons change from $-22, -5.8, -40$ and $-62$ MeV to $-12.2, 4.6, -30.4$ and $-51.4$ MeV, respectively.
  At baryon density $4\rho_0$, these values  change to 
  $-135, -88, -221$ and $-271$ MeV at $k = 0$ and  $-68.4, -21, -154.7$ and $-204.7$ MeV at $k = 500$ MeV.

In the present calculations of mass modifications of kaons and antikaons in dense resonance matter at finite temperature we used the chiral SU(3) model under mean field approximation.
The properties of antikaons in the nuclear matter  have been studied in the literature using the couple channel approach also
\cite{ramos2000properties,Tolos:2000fj,Tolos:2006ny,Tolos:2008di,Lutz:2007bh}.
In the coupled channel, the in-medium properties of antikaons are calculated taking into account the dynamics of $\Lambda(1405)$ resonance. In the free space the resonance $\Lambda(1405)$ is generated below the $\bar{K}N$ threshold 
considering the coupling of $\bar{K}N$ to $\pi\Sigma$  in isospin $I = 0$ channel\cite{Siegel:1988rq,Borasoy:2004kk,Borasoy:2006sr}.
At finite density of nuclear matter, the position of $\Lambda(1405)$ is shifted
to higher value of center of  mass energy due to Pauli blocking effects as has been studied using separable potential approach in Ref. \cite{Koch:1994mj}. When the calculations are performed self-consistently in the coupled channel approach, taking into account the in-medium self-energy and Pauli blocking simultaneously, the position of $\Lambda(1405)$ again shifts to lower baryon densities \cite{Lutz:1997wt}. The main objective of our present work was to investigate the in-medium masses and optical potentials of kaons and antikaons in the presence of decuplet baryons at finite baryon densities and temperatures and coupled channel dynamics has not been considered.
Apart from the chiral SU(3) model, the properties of kaons and antikaons have also been investigated using other relativistic models \cite{Friesen:2018ojv,Xu:2015mha}, QMC model \cite{Yue:2008tp,Tsushima:1997cu} and QCD sum rule approach \cite{Bozkir:2022lyk,Song:2018plu} without coupled channel dynamics.

Also, the study of antikaons at finite baryon densities is also relevant from neutron stars point of view due to the phenomenon of antikaon condensation.
Due to attractive  mass shift of antikaons as a function of baryon density,  $K^{-}$ mesons may play  the role of electrons to maintain the charge neutrality once the necessary $\beta-$equilibrium condition is satisfied. The phenomenon of antikaon condensation in presence of $\Delta$ baryon resonances has been investigated using the relativistic mean field model in 
Ref.~\cite{Thapa:2021kfo,Ma:2022knr} . 
In our future work, we shall extend our present work to explore physics of compact stars considering antikaon condensation in presence of decuplet baryons within the chiral SU(3) model.

\par
%Kaplan and Nelson suggest that, at low densities, there is attractive optical potential for antikaons. However, experimental data indicates, where the impulse approximation is applicable, that the $\bar K$ optical potential is a positive, demonstrating a repulsive interaction. This discrepancy arises because the $\bar K N$ scattering amplitude calculation does not account for the presence of $\Lambda$ (1405) resonance state, occurring in $\bar K$ proton-channel slightly below the threshold \cite{Koch:1994mj}, unlike in the coupled channel approach. The studies on kaon and antikaon properties within a medium, utilizing the mean-field approach, serve as an alternative method employed by various research groups \cite{Menezes:2005ic,Banik:2000dx,Glendenning:1998zx,kumar2020phi}. This approach is particularly useful for examining optical potentials at high baryonic densities, which may be present in neutron star cores and potentially generated in the CBM experiment of the FAIR project. In the coupled channel approach, an attractive potential arises even at low densities by incorporating the medium effects through Pauli blocking and self-consistent consideration of antikoans self-energy \cite{Ramos:1999ku,Tolos:2000fj,Tolos:2006ny,Tolos:2008di,Lutz:2007bh}.     
\section{Summary and conclusion}
\label{summary}
% The different potentials of kaons and antikaons may be especially significant for neutron-rich heavy-ion beams in experiments at the CBM facility at GSI-FAIR in Germany, as well as in future experiments at the proposed Rare Isotope Accelerator (RIA) laboratory in the United States.           may be used in future studies to learn more about the properties of the hadronic phase. The observed
To summarize, in the present work, using the chiral SU(3) mean field model, we calculated the in-medium masses and optical potentials of $K (K^+, K^0)$  and $\bar{K} (K^-, \bar{^0})$ mesons at finite baryon density and temperature in isospin asymmetric hadronic medium consisting of nucleons, hyperons and decuplet baryons. 
Employing the mean-field approximation, we evaluated  the properties of octet and decuplet baryons within the chiral SU(3) model through the exchange of scalar fields $\sigma, \zeta$ and $\delta$ and the vector fields $\omega, \rho$ and $\phi$. 
The interactions of kaons and antikaons with baryons and  the scalar fields ($\sigma, \zeta$ and $\delta$) are described in terms of 
 vectorial interaction term, scalar meson exchange term and the range terms.
 The in-medium masses of kaons and antikaons
 are modified significantly at finite temperature, when the decuplet baryons are included along with nucleons and hyperons in the equation of motions of scalar and vector fields to obtain the density and temperature dependent effects. We observed that in the presence of resonance baryons, at finite baryon density, the masses of kaons and antikaons decreases more significantly with increase in temperature of the medium from zero to 100 and 150 MeV. 
 However, without decuplet baryons, the masses of kaons and antikaons were observed to increase in the nuclear and hyperonic medium, as temperature is increased from zero to finite value \cite{Mishra:2004te, Mishra:2006wy, Mishra:2008kg,Mishra:2008dj,kumar2020phi}.
 This implies that the presence of decuplet baryons at finite temperature and density  of the medium contribute to the
 attractive interactions. 
 In the asymmetric non-strange medium with $I_a = 0.3$, including nucleons and $\Delta$ baryons, at $\rho_B = 4\rho_0$, the effective masses
 of $K^{+}, K^{0}, K^{-}$ and $\bar{K^0}$ mesons decreases by $49$, $50$, $90$ and $80$ MeV, as temperature is changed from zero to $100$ MeV.
 Including all the decuplet baryons along with octet of nucleons and hyperons, the above values  change to $15$, $34$, $154$ and $114$ MeV, respectively. The investigation of the optical potentials of kaons and antikaons reveals that adding more strange particles in the isospin-dependent medium has more appreciable modification in their values. 
 
 %baryons in an asymmetric hyperonic medium, it is observed that the tendency to restore chiral symmetry is enhanced, and the strangeness fraction  significantly influences the scalar fields and effective masses of $K^+$, $K^0$, $K^-$, and $\bar K^0$ compared to an isospin asymmetric. The antikaon masses exhibit the highest sensitivity to temperature compared to $I_a$ ans $f_s$, whereas, masses of kaons show more dependence on strangeness fraction. 
% Among the variation in $I_a$, $f_s$ and $T$, temeprature variation has more pronounced effect on the masses of antikaons compared to kaons.
%The investigation of the optical potentials of kaons and antikaons reveals that adding more strange particles in the isospin-dependent medium has more appreciable modification in their values. 

In-medium masses and optical potentials of kaons and antikaons may be used as input in various transport models for understanding 
the impact on different experimental observables \cite{Feng:2015vra,Steinheimer:2024eha}. For example, as discussed in Ref. \cite{Feng:2015vra} using
Lanzhou quantum molecular
dynamics (LQMD) transport model, the in-medium optical potentials and isospin effects are important for understanding the production rate of particles and their distribution at freeze out in phase space. The isospin effects on the optical potentials will be useful for the ratios $K^0/K^+$
and $K^-/\bar{K^0}$.
The in-medium masses of kaons and antikaons may also be used as input in exploring the $\phi$ meson production in heavy-ion collisions \cite{Song:2022jcj}.
Future experiments like the CBM and PANDA experiments of FAIR project at GSI, Germany \cite{klochkov2021compressed,Destefanis:2013xpa,senger2012compressed} and
Nuclotron-based Ion Collider Facility (NICA) at Dubna, Russia \cite{Kekelidze:2017ghu,Kekelidze:2017tgp},
aim to explore how the hadrons properties are modified at finite baryon density. Our present theoretical calculations may be used in future studies to gain further insight into the properties of the hadronic phase and the physics of heavy ion collisions. 
 \section*{Acknowledgment}

The authors sincerely acknowledge the support for this work from the Ministry of Science and Human Resources (MHRD), Government of India, through an Institute fellowship under the National Institute of Technology Jalandhar.
Arvind Kumar sincerely acknowledge 
Anusandhan National Research Foundation (ANRF),
Government of India for funding of the research project under the
Science and Engineering
Research Board-Core Research Grant (SERB-CRG) scheme (File No. CRG/2023/000557).
% for financial support.

\par

\bibliographystyle{elsarticle-num} 
%\biboptions{numbers,sortcompress}

\bibliography{Ref}

%\begin{thebibliography}{100}
%\bibitem{Azizi:2016hbr}
%K.~Azizi, N.~Er and H.~Sundu,
%%``Impact of finite density on spectroscopic parameters of decuplet baryons,''
%Phys. Rev. D \textbf{94}, 114002 (2016).
%\end{thebibliography}

%\bibliographystyle{unsrt}
%\bibliography{ref}
%\input{refe}
% \end{linenumbers}
\end{document}